\documentclass[singlespacing]{elsart}

\usepackage{graphicx}

\usepackage{amssymb,amsbsy}

\begin{document}



\begin{frontmatter}



\title{Dynamic Subgrid Scale Modeling of Turbulent Flows using Lattice-Boltzmann Method}

\author[label1,label2]{Kannan N. Premnath\corauthref{cor1}}
\ead{nandha@metah.com}
\author[label1]{Martin J. Pattison}
\author[label1,label2,label3,label4]{Sanjoy Banerjee}
\address[label1]{MetaHeuristics LLC, 3944 State Street, Suite 350, Santa Barbara, CA 93105}
\address[label2]{Department of Chemical Engineering, University of California Santa Barbara, Santa Barbara, CA 93106}
\address[label3]{Department of Mechanical Engineering, University of California Santa Barbara, Santa Barbara, CA 93106}
\address[label4]{Bren School of Environmental Science and Management, University of California Santa Barbara, Santa Barbara, CA 93106}
\corauth[cor1]{Corresponding author.}

\author{}

\begin{abstract}
In this paper, we discuss the incorporation of dynamic subgrid scale (SGS) models in the lattice-Boltzmann method (LBM)
for large-eddy simulation (LES) of turbulent flows. The use of a dynamic procedure, which involves sampling or
test-filtering of super-grid turbulence dynamics and subsequent use of scale-invariance for two levels, circumvents the need for empiricism in determining the magnitude of the model coefficient of the SGS models. We employ the multiple relaxation times (MRT) formulation of LBM with a forcing term, which has improved physical fidelity and numerical stability achieved by proper separation of relaxation time scales of hydrodynamic and non-hydrodynamic modes, for simulation of the grid-filtered dynamics of large-eddies. The dynamic procedure is illustrated for use with the common Smagorinsky eddy-viscosity SGS model, and incorporated in the LBM kinetic approach through effective relaxation time scales. The strain rate tensor in the SGS model is locally computed by means of non-equilibrium moments of the MRT-LBM. We also discuss proper sampling techniques or test-filters that facilitate implementation of dynamic models in the LBM. For accommodating variable resolutions,
we employ conservative, locally refined grids in this framework. As examples, we consider the canonical anisotropic and inhomogeneous turbulent flow problem, i.e. fully developed turbulent channel flow at two different shear Reynolds numbers $Re_{*}$ of $180$ and $395$. The approach is able to automatically and self-consistently compute the values of the Smagorinsky coefficient, $C_S$. In particular, the computed value in the outer or bulk flow region, where turbulence is generally more isotropic, is about $0.155$ (or the model coefficient $C=C_S^2=0.024$) which is in good agreement with prior data. It is also shown that the model coefficient becomes smaller and approaches towards zero near walls, reflecting the dampening of turbulent length scales near walls. The computed turbulence statistics at these Reynolds numbers are also in good agreement with prior data. The paper also discusses a procedure for incorporation of more general scale-similarity based SGS stress models.
\end{abstract}

\end{frontmatter}

\section{Introduction} \label{section_introduction}
Turbulence in fluids is ubiquitous in nature and technological systems and
represents one of the most challenging aspects in fluid mechanics.
The difficulty stems from the inherent presence of many scales that are
generally inseparable, among many other factors. Nevertheless, considerable
progress has been made over the years towards more fundamental
physical understanding of turbulence phenomena through measurements, statistical
phenomenological theories, modeling and computation~\cite{frisch95,pope00,lesieur07}.

In the case of their modeling and computation, different approaches, depending
on the desired degree of detail, have been devised. One approach is to employ
Reynolds-averaged equations of fluid motion representing the evolution of mean turbulence
field. In this framework, the averaged turbulent fluctuations -- the Reynolds stresses --
need to be modeled. Various turbulence models have been developed for this purpose, including
the well-known $k-\epsilon$ and Reynolds stress transport equations. A major limitation of
this approach is that Reynolds-averaged models need to represent physics over a wide range of scales.
While small scale turbulence tends be somewhat more universal, large scale turbulent motions
are strongly problem dependent. Thus, it is unrealistic to expect the Reynolds-averaged
models to accommodate and represent large-scale behavior of different classes of turbulent flows
in the same manner without resorting to considerable empiricism.

On the other hand, direct numerical simulation (DNS) approach resolves all spatio-temporal scales up to dissipation
scales without the use of models, and can thus predict all possible
motions, structural and statistical features due to turbulence with high
fidelity. Due to its high computational cost, its application is restricted to relatively low Reynolds numbers.
On the other hand, in view of this, it is more pragmatic to only compute the eddying motions of fluids
with length scales greater than the grid size and model unresolved subgrid scales (SGS), which comprises the
large eddy simulation (LES) approach. The SGS scales, which mainly account for dissipation of turbulence,
are generally more isotropic in nature and relatively independent of the resolved large-scale fluid dynamics.
Thus, LES represents a compromise with reduced empiricism in contrast to Reynolds-averaged models, but with reduced
computational cost in comparison with DNS.

In LES approach, the Smagorinsky eddy-viscosity model (1963)~\cite{smagorinsky63} is one of the simplest SGS models. Based on the assumption that the small scales are in equilibrium, they dissipate all the energy transferred from large
scales instantaneously and completely. This dissipation mechanism is modeled by an eddy viscosity, which is represented
by the product of the square of a length scale, which is often the grid size, and the inverse of a time scale, which is given by the local fluid shear rate, with a proportionality coefficient or ``constant". The values of this proportionality ``constant" were first determined by Lilly (1966)~\cite{lilly66}  and Deardorff (1970)~\cite{deardorff70} by analysis for isotropic
turbulence and turbulent channel flow, respectively, and one of the first successful LES applications is by Schumann (1975)~\cite{schumann75}. A more theoretical basis for LES was first provided by Leonard (1974)~\cite{leonard74} through the introduction of formal filtering procedures on fluctuating turbulence fields.

In wall-bounded flows, turbulence becomes anisotropic close to walls, with its length scales becoming progressively smaller and vanishing at the wall. Thus, with the use of a finite grid size as the length scale in the SGS model, the proportionality coefficient should decrease and approach to zero towards the wall to account for near-wall turbulence anisotropy and the model coefficient is actually not a ``constant". Often, an ad hoc damping function~\cite{vanDriest56} is used for this purpose~\cite{moin82}. Moreover, in transitional flows, it is found the use of the ``constant" Smagorinsky SGS model in LES can cause excessive damping of the resolved structures~\cite{piomelli90} and can be alleviated by an additional ad hoc intermittency function. The need to \emph{a priori} specify the values of the proportionality coefficient and their variation, for example, in the presence of turbulence anisotropy or laminar-to-turbulence transition comprises the empirical elements of the LES approach.

Significant progress in the LES approach was made by the introduction of dynamic procedures for local computation of
the coefficient(s) in the SGS models by Germano and co-workers~\cite{germano91,germano92}. By obtaining information of the turbulence field at scales larger than the grid size -- the``test" filter scale, and relating them with the resolved turbulence at grid scale through scale-invariance, the value of the model coefficient can be computed dynamically as the computation progresses. Thus, dynamic procedures avoid the need to \emph{a priori} specify the model coefficient and determine them based on the local fluctuating turbulence field. Moreover, unlike the constant Smagorinsky SGS model, it does not rule out the possibility of backscattering of energy from smaller scales. It is found that the use of dynamic procedure makes LES susceptible to numerical instabilities, which can be substantially alleviated by employing a least squares technique~\cite{lilly92} in conjunction with the use of averaging to compute the coefficient.

When the dynamic procedure is applied to the Smagorinsky SGS model, it is often simply referred to as the dynamic Smagorinsky model (DSM). Other SGS models include the use of a scale-similarity
stress tensor combined with the Smagorinksy model -- the mixed SGS model~\cite{bardina80} -- to provide improved predictions for certain classes of inhomogeneous flows. Dynamic procedures have been applied with much success to mixed SGS models, resulting in different variants such as the dynamic mixed model (DMM)~\cite{zang93} and the dynamic two-parameter model (DTM)~\cite{salvetti95}. The remarkable progress made in the development of LES
as a technique for turbulence simulation and their applications have been discussed in detail in the reviews by Piomelli (1999)~\cite{piomelli99}, Meneveau and Katz (2000)~\cite{meneveau00} and Pope (2004)~\cite{pope04}, and in the monographs of Galperin and Orszag (1993)~\cite{galperin93} and Sagaut (2003)~\cite{sagaut03}.

In this paper, we propose to incorporate dynamic procedure in the lattice Boltzmann method (LBM) for LES. In recent years, the LBM, based on minimal discrete kinetic models, has emerged as an alternative and accurate computational approach for fluid mechanics problems~\cite{chen98,succi01}. It involves the solution of the lattice-Boltzmann equation (LBE) that represents the changes in the evolution of the distribution of particle populations due to their advection, represented as a free-flight process, and collision, described as a relaxation process, on a lattice. When the lattice, which represents the discrete directions for particle propagation, satisfies sufficient rotational symmetries, the averaged LBE asymptotically recovers the weakly compressible Navier-Stokes equations (NSE).

Though it evolved as a computationally efficient form of lattice gas
cellular automata~\cite{higuera89}, it was well established about a
decade ago that the LBE is actually a much simplified form of the
Boltzmann equation~\cite{he97,abe97}. As a result, several previous
results in discrete kinetic theory could be directly applied to the
LBE. This lead to, for example, improved physical modeling in
various situations, such as multiphase flows~\cite{luo00,he02} and
multicomponent flows~\cite{asinari06}, and in an asymptotic theory
suitable for rigorous numerical analysis~\cite{junk05}. As a result of features of the stream-and-collide
procedure of the LBE such as the algorithmic simplicity, amenability
to parallelization with near-linear scalability, and its ability to
represent complex boundary conditions and incorporate physical
models more naturally, it has rapidly found a wide range of
applications~\cite{ladd01,succi02,yu03,nourgaliev03}.

Discrete kinetic theory, and in particular the LBM, has recently been employed to
\emph{a priori} derive turbulence models based on a mean-field
approximation~\cite{chen98a,ansumali04,chen04} and lattice kinetic
turbulence models have found much success~\cite{chen03}.
It has now been well established that LBM is a reliable and accurate method for
direct numerical simulation (DNS) of various benchmark turbulent
flow problems~\cite{martinez94,amati97,yu05a,yu05b,lammers06}.
On the other hand, in earlier work, the use of LBM as a LES tool with
``constant" Smagorinsky model to represent SGS effects was also
proposed~\cite{eggels96,hou96}, which has more recently found applications
for flows in different configurations and
physical conditions~\cite{lu02,krafczyk03,yu05c,yu06}.

A commonly used form of the LBM employs a single relaxation time
(SRT) model~\cite{bhatnagar54} to represent the effect of particle
collisions, in which particle distributions relax to their local
equilibrium at a rate determined by a single
parameter~\cite{qian92,chen92}. On the other hand, an equivalent
representation of distribution functions is in terms of their
moments, such as various hydrodynamic fields including density, mass
flux, and stress tensor. The relaxation process due to collisions
can more naturally be described in terms of a space spanned by such
moments, which can, in general relax at different rates. This forms
the basis of the generalized lattice-Boltzmann equation (GLBE) based
on multiple relaxation times
(MRT)~\cite{dhumieres92,lallemand00,dhumieres02}.

By carefully separating the time scales of various hydrodynamic and kinetic modes
through a linear stability analysis, the numerical stability of the
GLBE or MRT-LBE can be significantly improved when compared with the
SRT-LBE, particularly for more demanding problems at high Reynolds
numbers~\cite{lallemand00}. MRT-LBE has also been extended for
multiphase flows with superior stability
characteristics~\cite{mccracken05,premnath06,premnath05a,mccracken05a},
and, more recently, for magnetohydrodynamic
problems~\cite{pattison07}. It has also been used for LES of a class
of turbulent flows~\cite{krafczyk03,yu06}. Recently, we have extended the GLBE with forcing terms for
wall-bounded turbulent flows~\cite{premnath08}, which is a generalization of the approach introduced earlier by
Yu \emph{et al}.~\cite{yu06}. These forcing terms, determined
to avoid discrete lattice artifacts, represent the effect of external forces and are considered in
natural moment space of GLBE. In this work, we represented the
SGS effects through the relaxation times of the hydrodynamic moments using
the ``constant" Smagorinsky eddy-viscosity model~\cite{smagorinsky63}, which is modified by the van Driest
damping function~\cite{vanDriest56} in the near wall region.

One of the main objectives of this paper is to develop and implement a framework for
application of dynamic procedure in the lattice Boltzmann method (LBM) for LES of
inhomogeneous and anisotropic turbulent flows. In view of the previous discussions,
we plan to employ GLBE based on multiple relaxation times as the basis for this purpose.
We will present the incorporation of the dynamic procedure for the Smagorinsky SGS model, i.e.
DSM, in LBM. Particular attention is paid to performing discrete filtering operations at super
lattice grid or test-filter levels, which will be discussed.
To accommodate efficient simulation of varying turbulent length scales in the near-wall region,
variable resolution strategies are required. In this regard, we employ conservative, locally refined multiblock grids
that allow variable resolution. The approach will be tested for the fully-developed turbulent channel flow problem at two different shear Reynolds
numbers $Re_{*}$ of $180$ and $395$, by comparing the turbulent statistics with prior data based on
direct numerical simulations (DNS) and measurements. The higher Reynolds number case, to the best of our
knowledge, is considered for the first time in the LBM for LES of turbulent channel flows. The incorporation of DSM in the GLBE is a first step in the application of dynamic procedure. More general SGS models, involving scale-similarity stresses, such as DMM and DTM can also be introduced in the GLBE, which will also be briefly described in an appendix of this paper, and their implementation will be a subject of future work.

The organization of this paper is as follows: In Sec.~\ref{subsec:glbe}, we will discuss the generalized lattice-Boltzmann equation (GLBE) with forcing term. The use of conservative locally refined multiblock grids
will be presented in Sec.~\ref{sec:multiblockturbchannelflow}.
Section~\ref{sec:dynamicSGS} will discuss the dynamic SGS modeling with DSM in GLBE framework. Then, Sec.~\ref{sec:results} will present the LES
results for the canonical turbulent channel flow problem at two different Reynolds numbers. Finally, the summary
and conclusions of this paper will be presented in Sec.~\ref{sec:summary}.

\section{\label{subsec:glbe}Generalized Lattice Boltzmann Equation with Forcing Term}
We shall now discuss the computational procedure based on
generalized lattice-Boltzmann equation with a forcing term, which is
supplemented by Smagorinsky SGS model with a variable coefficient, obtained by means of
a dynamic procedure (see next section). For brevity,
we will present only the major elements of the approach, while the
details can be found in
Refs.~\cite{dhumieres02,yu06,premnath06,premnath08}.

\begin{figure}
\includegraphics[width = 80mm,viewport=185 150 540
460,clip]{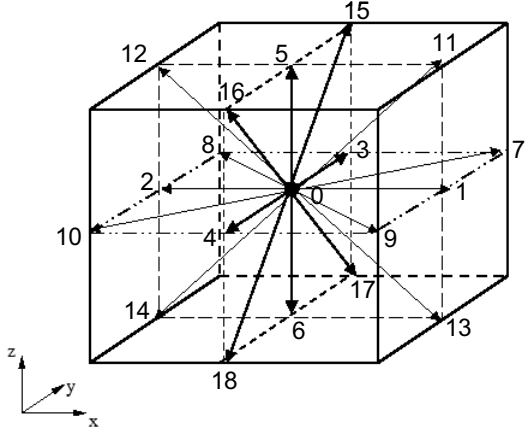}
\caption{\label{fig:d3q19} Schematic illustration of the
three-dimensional, nineteen velocity (D3Q19) model.}
\end{figure}

The lattice-Boltzmann method computes the evolution of distribution
functions as they move and collide on a lattice grid. The collision
process consider their relaxation to their local equilibrium values,
and the streaming process describes their movement along the
characteristics directions given by a discrete particle velocity
space represented by a lattice. Figure~\ref{fig:d3q19} represents
the three-dimensional, nineteen particle velocity (D3Q19) lattice
model employed in this paper.
The particle velocity $\overrightarrow{e_{\alpha}}$ corresponding to this
model may be written as:
\begin{equation}
\overrightarrow{e_{\alpha}} = \left\{\begin{array}{ll}
   {(0,0,0)}&{ \alpha=0}\\
   {(\pm 1,0,0),(0,\pm 1,0),(0,0,\pm 1)}&{ \alpha=1,\cdots,6}\\
   {(\pm 1,\pm 1,0),(\pm 1,0,\pm 1),(0,\pm 1,\pm 1)}&{ \alpha=7,\cdots,18}
\end{array} \right.
\label{eq:velocityd3q19}
\end{equation}
The GLBE computes collision in moment space, while the streaming
process is performed in the usual particle velocity
space~\cite{dhumieres02}. The GLBE with forcing
term~\cite{premnath08} also computes the forcing term, which
represents the effect of external forces as a second-order accurate
time-discretization, in moment space. We use the following notation
in our description of the procedure below: In \emph{particle
velocity space}, the local distribution function $\mathbf{f}$, its
local equilibrium distribution $\mathbf{f}^{eq}$, and the source
terms due to external forces $\mathbf{S}$ may be written as the
following column vectors: $\mathbf{f}=\left[
f_0,f_1,f_2,\ldots,f_{18} \right]^{\dag}$, $\mathbf{f}^{eq}=\left[
f_0^{eq},f_1^{eq},f_2^{eq},\ldots,f_{18}^{eq} \right]^{\dag}$, and $
\mathbf{S}=\left[ S_0,S_1,S_2,\ldots,S_{18} \right]^{\dag}$. Here,
the superscript $\dag$ represents the transpose operator.

The moments $\mathbf{\widehat{f}}$ are related to the distribution
function $\mathbf{f}$ through the relation
$\mathbf{\widehat{f}}=\mathcal{T}\mathbf{f}$ where $\mathcal{T}$ is
the transformation matrix. Here, and in the following, the ``hat"
represents the moment space. The transformation matrix $\mathcal{T}$
is constructed such that the collision matrix in moment space
$\widehat{\Lambda}$ is a diagonal matrix through
$\widehat{\Lambda}=\mathcal{T}\Lambda\mathcal{T}^{-1}$, where
$\Lambda$ is the collision matrix in particle velocity space. The
elements of $\mathcal{T}$ are obtained in a suitable orthogonal
basis as combinations of monomials of the Cartesian components of
the particle velocity $\overrightarrow{e_{\alpha}}$ through the
standard Gram-Schmidt procedure, which are provided by
d'Humi{\`e}res \emph{et al}. (2002)~\cite{dhumieres02}. Similarly,
the equilibrium moments and the source terms in moment space may be
obtained through the transformation
$\mathbf{\widehat{f}}^{eq}=\mathcal{T}\mathbf{f}^{eq}$,
$\mathbf{\widehat{S}}=\mathcal{T}\mathbf{S}$. The components of
moment-projections of these quantities are:
$\mathbf{\widehat{f}}=\left[
\widehat{f}_0,\widehat{f}_1,\widehat{f}_2,\ldots,\widehat{f}_{18}
\right]^{\dag}$, $\mathbf{\widehat{f}}^{eq}=\left[
\widehat{f}_0^{eq},\widehat{f}_1^{eq},\widehat{f}_2^{eq},\ldots,\widehat{f}_{18}^{eq}
\right]^{\dag}$, and $ \mathbf{\widehat{S}}=\left[
\widehat{S}_0,\widehat{S}_1,\widehat{S}_2,\ldots,\widehat{S}_{18}
\right]^{\dag}$. These are provided in
Appendix~\ref{app:momentcomponents}.

The solution of the GLBE with forcing term can be written in terms
of the following ``effective" collision and streaming steps,
respectively:
\begin{equation}
\mathbf{\widetilde{f}}(\overrightarrow{x},t)=\mathbf{f}(\overrightarrow{x},t)+\boldsymbol{\varpi}(\overrightarrow{x},t),
\label{eq:postcollision}
\end{equation}
and
\begin{equation}
f_{\alpha}(\overrightarrow{x}+\overrightarrow{e}_{\alpha}\delta_t,t+\delta_t)=\widetilde{f}_{\alpha}(\overrightarrow{x},t),
\label{eq:streaming}
\end{equation}
where the distribution function
$\mathbf{f}=\{f_{\alpha}\}_{\alpha=0,1,\ldots,18}$ is updated due to
``effective" collision resulting in the post-collision distribution
function
$\mathbf{\widetilde{f}}=\{\widetilde{f}_{\alpha}\}_{\alpha=0,1,\ldots,18}$
before being shifted along the characteristic directions during
streaming step. $\boldsymbol{\varpi}$ represents the change in
distribution function due to collisions as a relaxation process and
external forces, and following Premnath \emph{et al}.
(2008)~\cite{premnath08} it can written as
\begin{equation}
\boldsymbol{\varpi}(\overrightarrow{x},t)= \mathcal{T}^{-1}\left[
-\widehat{\Lambda}\left(\mathbf{\widehat{f}}-\mathbf{\widehat{f}}^{eq}
\right)_{(\overrightarrow{x},t)}+\left(
\mathcal{I}-\frac{1}{2}\widehat{\Lambda}
\right)\mathbf{\widehat{S}}_{(\overrightarrow{x},t)} \right],
\label{eq:relax_term}
\end{equation}
where $\mathcal{I}$ is the identity matrix and $
\widehat{\Lambda}=diag(s_0,s_1,\ldots,s_{18})$ is the diagonal
collision matrix in moment space. Now, some of the relaxation times
$s_{\alpha}$ in the collision matrix, i.e. those corresponding to
hydrodynamic modes can be related to the transport coefficients and
modulated by eddy-viscosity due to SGS model (discussed below) as
follows~\cite{dhumieres02,yu06,premnath08}:
$s_{1}^{-1}=\frac{9}{2}\zeta+\frac{1}{2}$, where $\zeta$ is the
molecular bulk viscosity, and
$s_{9}=s_{11}=s_{13}=s_{14}=s_{15}=s_{\nu}$, where $ s_{\nu}^{-1}=
3\nu+\frac{1}{2}= 3(\nu_0+\nu_t)+\frac{1}{2}$, with $\nu_0$ being
molecular shear viscosity and $\nu_t$ the eddy-viscosity determined
from the dynamic SGS model discussed in the next section.
The rest of the relaxation parameters, i.e. for the kinetic modes,
can be chosen through a von Neumann stability
analysis of the linearized GLBE~\cite{dhumieres02}: $s_1=1.19$,
$s_2=s_{10}=s_{12}=1.4$, $s_4=s_6=s_8=1.2$ and
$s_{16}=s_{17}=s_{18}=1.98$.

Once the distribution function is known, the hydrodynamic fields,
i.e., the density $\rho$, velocity $\overrightarrow{u}$ and pressure
$p$ can be obtained as follows:
\begin{equation}
\rho =\sum_{\alpha=0}^{18} f_{\alpha}, \quad
\overrightarrow{j}\equiv \rho\overrightarrow{u}
=\sum_{\alpha=0}^{18}
f_{\alpha}\overrightarrow{e}_{\alpha}+\frac{1}{2}\overrightarrow{F}\delta_t,
\quad p=c_s^2\rho,
\label{eq:macrofieldvar}
\end{equation}
where, $c_s=c/\sqrt{3}$ with $c=\delta_x/\delta_t$ being the
particle speed, and $\delta_x$ and $\delta_t$ are the lattice
spacing and time step, respectively. It can be readily shown that when a
multiscale analysis based on the Chapman-Enskog expansion~\cite{chapman64,premnath06}
is applied to the GLBE with relaxation times scales augmented by an eddy-viscosity,
it recovers the ``grid filtered" weakly compressible Navier-Stokes
equations with external forces. That is, the density, momentum and pressure obtained
from Eq.~(\ref{eq:macrofieldvar}) are grid-filtered quantities: $\rho\rightarrow \overline{\rho}$,$\rho\overrightarrow{u}\rightarrow \overline{\rho}\overrightarrow{\overline{u}}$ and $p\rightarrow \overline{p}$, where the `overbar' represents grid-filter. It may be noted that the macrodynamical equations can alternatively be derived through
an asymptotic analysis under a diffusive scaling~\cite{junk05}.

The computational procedure for the solution of the GLBE with
forcing term is optimized by fully exploiting the special properties
of the transformation matrix $\mathcal{T}$: these include its
orthogonality, entries with many zero elements, and entries with
many common elements that are integers, which are used to form the
most common sub-expressions for transformation between spaces in
avoiding direct matrix multiplications~\cite{dhumieres02}. For
details, we refer the reader to Ref.~\cite{yu06,premnath08}. As a result
of such optimization, the additional computational overhead when
GLBE is used in lieu of the popular SRT-LBE is small, typically
between $15\%-30\%$, but with much enhanced numerical stability that
allows maintaining solution fidelity on coarser grids and also in
simulating flows at higher Reynolds numbers.

\section{\label{sec:multiblockturbchannelflow}Conservative Local Grid Refinement Based on Multiblock Approach}
Closer to a wall, length scales are very small, requiring a fine
grid to adequately resolve turbulent structures. Use of grid fine
enough to resolve the wall regions throughout the domain can entail
significant computational cost, and this can be mitigated by
introducing coarser grids farther from the wall, where turbulent
length scales are larger. One approach is to consider using
continuously varying grid resolutions, using a
interpolated-supplemented LBM~\cite{he96} that effectively decouples
particle velocity space represented by the lattice and the
computational grid. However, it is well known that interpolation
could introduce significant numerical dissipation, see for
e.g.~\cite{lallemand00}, which could severely affect the accuracy of
solutions involving turbulent fluctuations. Thus, we consider locally embedded grid
refinement approaches, and in particular their conservative
versions~\cite{chen06,rohde06} that preserve mass and momentum
conservation. Similar zonal embedded approaches have been
successfully employed in computational approaches based on the
solution of filtered NSE for LES of turbulent
flows~\cite{kravchenko96}.

\begin{figure}
\includegraphics[width = 120mm,angle=90]{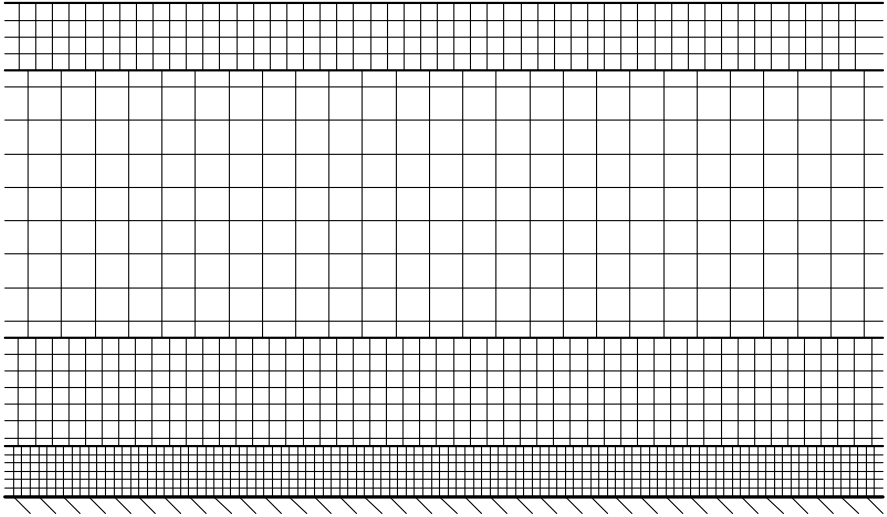}
\caption{\label{fig:multiblockgrid2} Schematic illustration of conservative, locally refined multiblock gridding strategy for
LES of turbulent channel flow using GLBE with dynamic SGS model.}
\end{figure}

Figure~\ref{fig:multiblockgrid2} shows a schematic of such a
multiblock approach in which a fine cubic lattice grid is used close
to the bottom wall surfaces and a coarser one, again cubic in shape, farther out. Similarly,
finer grids are employed near top free-slip surfaces.
In order to facilitate the exchange of information at the interface
between the grids, the spacing of the nodes changes by an integer
factor, in this case two.  As well as using different grid sizes,
the two regions use different time steps (time step being
proportional to grid size), and the computational cost required per
unit volume is thus reduced by a factor of 16 in the coarse grid.
Figure~\ref{fig:multiblockgrid2} shows a staggered  grid
arrangement, in which nodes on the fine and coarse sides of the
interface are arranged in a manner that facilitates the imposition
of mass and momentum conservation. Different blocks communicate with
each other through the \emph{Coalesce} and the \emph{Explode} steps,
in addition to the standard \emph{stream-and-collide} procedure. The
details are provided in Refs.~\cite{chen06,rohde06} and we very
briefly present the essential elements in the following.

The \emph{Coalesce} procedure involves summing the particle populations
on the fine nodes to provide new incoming particle populations for
the corresponding coarse nodes. Similarly, the \emph{Explode} step
involves redistributing the populations on the coarse node to the
surrounding fine nodes. These grid~-~communicating steps used in the
multiblock approach presented in Ref.~\cite{chen06} were
incorporated in GLBE framework in this work. It may be noted that other
approaches to local grid refinement exist in the literature. In particular,
asymptotic analysis of the LBE shows that it converges to the incompressible NSE
when the ratio of the square of time step and
square of the grid spacing is a constant~\cite{junk05}. Local grid refinement based
on this consideration need to adapt the time step to vary as the square of the grid spacing.
However, in this work we use the grid refinement variant that considers
linear variation of the time step with the grid spacing for reasons of
relatively lower computational cost and the fact that it is already more developed and tested
for different problems~\cite{chen06,rohde06}.

\section{\label{sec:dynamicSGS} Subgrid Scale Eddy-Viscosity Modeling using Dynamic Procedure}
The main idea behind the application of dynamic procedure to SGS models~\cite{germano91}
is that the local value of the model coefficient of the eddy viscosity, representing eddy-viscosity
type subgrid scale effects, can be obtained from sampling the smallest super-grid or resolved scales,
which are generally referred to as the test-filtered scales, and assuming scale-invariance at these
two levels. If $\overline{\Delta}$ is the width of
the grid filter, which in the GLBE becomes $\overline{\Delta}=\delta_x$, where $\delta_x$ is the lattice
grid spacing, the flow information is sampled at a larger scale $\widetilde{\overline{\Delta}}$, the
test-filter width, i.e. $\widetilde{\overline{\Delta}}>\overline{\Delta}$, and generally $\widetilde{\overline{\Delta}}/\overline{\Delta}=2$~\cite{germano91}. The notation adapted here,
as before, and in the following is that `bar' refers to grid-filtered values and a `tilde' refers to
test-filtered values. The effect of subgrid scales is parameterized by an eddy viscosity relation
~\cite{smagorinsky63}
\begin{equation}
\nu_t=C\overline{\Delta}^2|\overline{S}|,
\label{eq:eddyviscosity}
\end{equation}
where $|\overline{S}|$ is the strain rate, given by $|\overline{S}|=\sqrt{2\overline{S}_{ij}\overline{S}_{ij}}$ and $C$ is the model coefficient. The grid-filtered strain rate $|\overline{S}|$ can be obtained by evaluating the corresponding strain rate tensor from the non-equilibrium moments as shown in Appendix~\ref{app:momentcomponents}. The model coefficient $C$ is obtained as follows.

The anisotropic part of the SGS stress at grid-filter scale $\tau_{ij}$ ($\tau_{ij}=\overline{u}_i\overline{u}_j-\overline{u_iu_j}$) and that at the test-filter scale $T_{ij}$ ($T_{ij}=\widetilde{\overline{u}}_i\widetilde{\overline{u}}_j-\widetilde{\overline{u_iu_j}}$), where $\overline{u_iu_j}$ and $\widetilde{\overline{u_iu_j}}$ are unknowns,
are modeled in terms of eddy viscosity and strain rates at corresponding scales, respectively, or by
invoking scale-invariance, as
\begin{equation}
\tau_{ij}-\frac{\delta_{ij}}{3}\tau_{kk}=-2C\overline{\Delta}^2|\overline{S}|\overline{S}_{ij}=-2\nu_t\overline{S}_{ij},
\label{eq:gridfilterstress}
\end{equation}
and
\begin{equation}
T_{ij}-\frac{\delta_{ij}}{3}T_{kk}=-2C\widetilde{\overline{\Delta}}^2|\widetilde{\overline{S}}|\widetilde{\overline{S}}_{ij}=-2\nu_t\widetilde{\overline{S}}_{ij},
\label{eq:testfilterstress}
\end{equation}
where $\overline{S}_{ij}$ and $\widetilde{\overline{S}}$ refers to the strain rate tensors at the grid- and test- filter levels, respectively. The respective magnitude of strain rates $|\overline{S}|$ and $|\widetilde{\overline{S}}|$ may be written as:

$|\overline{S}|=\sqrt{2\overline{S}_{ij}\overline{S}_{ij}}, \quad \overline{S}_{ij}=1/2\left(\partial_i \overline{u}_j+\partial_j \overline{u}_i\right)$,

$|\widetilde{\overline{S}}|=\sqrt{2\widetilde{\overline{S}}_{ij}\widetilde{\overline{S}}_{ij}}, \quad \widetilde{\overline{S}}_{ij}=1/2\left(\partial_i \widetilde{\overline{u}}_j+\partial_j \widetilde{\overline{u}}_i\right)$.

It may be noted that the grid-filtered strain rates can be obtained locally from the non-equilibrium moments, as mentioned above; the test-filtered strain rates can be determined directly from the gradient of test-filtered
velocity field, which can, in turn, be obtained by explicit application of test-filter on the grid-filtered velocity
field. The expressions for the test-filter that can be conveniently used in the GLBE framework will be discussed later.

The unknown SGS stress at each filter level can be related by the Germano identity~\cite{germano91}
\begin{equation}
L_{ij}=\widetilde{\overline{u}_i\overline{u}_j}-\widetilde{\overline{u}_i}\widetilde{\overline{u}_j}=T_{ij}-\widetilde{\tau}_{ij},
\label{eq:resolvedstress}
\end{equation}
where $L_{ij}$ is the resolved turbulent stress, and upon substituting Eqs.~(\ref{eq:gridfilterstress}) and~(\ref{eq:testfilterstress}) in Eq.~(\ref{eq:resolvedstress}), we get an
expression between $L_{ij}$ and other known variables with the model coefficient $C$ being the only
unknown
\begin{equation}
L_{ij}-\frac{\delta_{ij}}{3}L_{kk}=-2CM_{ij},
\label{eq:Germanoexpression}
\end{equation}
where
\begin{equation}
M_{ij}=\widetilde{\overline{\Delta}}^2|\widetilde{\overline{S}}|\widetilde{\overline{S}}_{ij}-\overline{\Delta}^2\widetilde{|\overline{S}|\overline{S}_{ij}}.
\label{eq:MStressexpression}
\end{equation}
We note that the computation of the first term in the $M_{ij}$ tensor involves explicit test-filtering and finite-differencing, while its second term can be obtained locally from non-equilibrium moments, as discussed above. Since, the tensor expressions, i.e. Eq.~(\ref{eq:Germanoexpression}) and~(\ref{eq:MStressexpression}) leads to five independent equations containing $C$, it can be obtained by a least square minimization approach as~\cite{lilly92}:
\begin{equation}
C=-\frac{1}{2}\frac{\left<L_{kl}M_{kl}\right>}{\left<M_{ij}M_{ij}\right>},
\label{eq:Ccoefficient}
\end{equation}
where the usual summation convention of the repeated indices is assumed and $<\cdot>$ represents spatial averaging in homogeneous directions or time averaging or both, depending on
the problem, so as to stabilize the computations. Also clipping is to be done when $\nu_t<0$, i.e. set $\nu_t=0$ in such cases. Once the eddy viscosity in Eq.~(\ref{eq:eddyviscosity}) has been determined from Eqs.~(\ref{eq:resolvedstress})-(\ref{eq:Ccoefficient}) it is added
to the molecular viscosity $\nu_0$ , obtained from the statement of the problem, to yield the total
viscosity $\nu$, i.e. $\nu=\nu_0+\nu_t$, which can then be used to compute the ``effective" relaxation times for
hydrodynamic moments for the GLBE discussed in Sec.~\ref{subsec:glbe}.

\subsection{\label{subsec:filters}Test-Filters}
Several approaches exist to obtain the test-filtered quantities from the grid-filtered
variables that can, in turn, be directly obtained from the solution of the GLBE. The approach considered in this paper is the discrete trapezoidal filter~\cite{zang93,najjar96}. Discrete trapezoidal filters are employed in this work as they naturally accommodate with the underlying cubic lattice grid structure and thereby allowing an efficient implementation. They involve applying the one-dimensional filter successively in each of the three coordinate directions as follows:
\begin{eqnarray}
\overline{\phi}_{(i,j,k)}^{*}&=&1/4\left(\overline{\phi}_{(i+1,j,k)}+2\overline{\phi}_{(i,j,k)}+\overline{\phi}_{(i-1,j,k)}\right),\nonumber\\
\overline{\phi}_{(i,j,k)}^{**}&=&1/4\left(\overline{\phi}_{(i,j+1,k)}^{*}+2\overline{\phi}_{(i,j,k)}^{*}+\overline{\phi}_{(i,j-1,k)}^{*}\right),\nonumber\\
\widetilde{\overline{\phi}}_{(i,j,k)}&=&1/4\left(\overline{\phi}_{(i,j,k+1)}^{**}+2\overline{\phi}_{(i,j,k)}^{**}+\overline{\phi}_{(i,j,k-1)}^{**}\right),\nonumber
\end{eqnarray}
to obtain the test-filtered value $\widetilde{\overline{\phi}}$ from $\overline{\phi}$ at $(i,j,k)$. Near walls, along the coordinate direction $k$, the test-filtered values at the top and bottom can be obtained as
$\widetilde{\overline{\phi}}_{(i,j,k)}=1/2\left(\overline{\phi}_{(i,j,k)}^{**}+\overline{\phi}_{(i,j,k+1)}^{**}\right)$
and $\widetilde{\overline{\phi}}_{(i,j,k)}=1/2\left(\overline{\phi}_{(i,j,k-1)}^{**}+\overline{\phi}_{(i,j,k)}^{**}\right)$, respectively.

Other types of test-filters specialized for lattice stencils that can also maintain isotropy may also be
possible, but presumably at the expense of additional computational overhead. The implementation of test-filters in the multiblock approach (see Sec.~\ref{sec:multiblockturbchannelflow}) leads to additional challenges in how best to deal with performing such filtering operations at the grid boundaries. In order to calculate the model coefficient $C$, it is necessary to use information from two grid spacings away; thus to compute $C$ at the first node on the coarse grid,
two ghost nodes are introduced. Velocities at these ghost nodes are
obtained by averaging the values at the surrounding fine nodes. Note that since the collision step
is not performed on the two fine layers adjacent to the grid boundary, it is not necessary to
introduce additional ghost nodes for the fine grid. For an efficient implementation, optimization strategies are considered in the incorporation of the dynamic procedure in the multiblock GLBE.

It may be noted that the dynamic procedure with other SGS models, such as the dynamic mixed model (DMM)~\cite{zang93} and the dynamic two-parameter model (DTM)~\cite{salvetti95} can be incorporated in the GLBE in a similar manner. The details of the incorporation of these models in the GLBE are presented in Appendix~\ref{app:dynamicmixedmodels}.

\section{\label{sec:results}Results and Discussion}
We will now present investigations of the LBM using the GLBE formulation with dynamic SGS modeling
for LES of wall-bounded turbulent flows. In earlier studies, the LBM has been validated as a direct
numerical simulations (DNS) approach for such problems~\cite{amati97,lammers06}. In the
present study, we will assess the performance of GLBE as a LES approach that incorporates SGS effects
using a dynamic procedure to compute the model coefficient, thereby avoiding empiricism,
on a relatively coarse grid with multiblock approach, while maintaining necessary near-wall resolutions.

\subsection{Turbulent Channel Flow at $Re_{*}=180$}
First, we consider fully-developed turbulent flow in an open channel with a shear Reynolds number of $Re_{*}=u_{*}H/\nu_{0}=180$, where $u_{*}$ is the shear velocity, $\nu_0$ is the molecular kinematic
viscosity and $H$ is the channel height. The shear velocity is related to the wall stress $\tau_{w}$ by $u_{*}=\sqrt{\tau_{w}/\rho}$. It represents an important canonical problem to test turbulence models and computational techniques, for which detailed direct numerical simulations (DNS) data are available for comparison~\cite{kim87,lam92}.

\subsubsection{Computational Conditions}
No-slip boundary is considered at the bottom, implemented using half-way
link bounce back approach~\cite{ladd94}, free-slip at the top, by means of specular reflection of distribution functions~\cite{succi01}, and periodic boundary conditions are applied in the streamwise, $x$ and
spanwise, $y$ directions. The computational domain is chosen with appropriate aspect ratios, viz., $5.7H$ and $2.85H$ in the
streamwise and spanwise directions, respectively. With these aspect ratios, a sufficient number of wall-layer
streaks are accommodated and end effects of two-point correlations are excluded, i.e.
the two-point velocity correlations in solutions are required to decay nearly to zero within half the
domain~\cite{moin98}.

For wall-bounded turbulent flows, it is
important to adequately resolve the near-wall, small-scale turbulent
structures. Grid spacing for such problems are generally referred in terms
of wall units (designated with a ``$+$" superscript) as
$\Delta^{+} = \Delta/\delta_{\nu}$, where $\Delta$ is the wall spacing
and  $\delta_{\nu}=\nu_{0}/u_{*}$ is the characteristic viscous length
scale. When the computations resolve the local dissipative or Kolmogorov length scale
$\eta=(\nu^3_{0}/\epsilon)^{1/4}$, i.e. $\Delta^{+}_{nw}\leq O(\eta^{+})$~\cite{moin98},
it can satisfy the near-wall resolution requirements.
In particular, it is generally recognized that $1.5\eta^{+}-2.0\eta^{+}$ represents
the upper limit of grid-spacing, above which the small scale turbulent motions in
bounded flows are not well resolved. It can be shown by simple
arguments that $\eta^{+}\approx 1.5-2.0$ at the wall and that
$\eta^{+}$ increases with increasing distance from the
wall~\cite{pope00}.

To account for resolution requirements, we discretize the computational domain in terms of
two blocks with different resolutions, viz., a fine block near the wall and a coarse block,
whose grid spacings are twice that of the fine-block in each direction, in the
bulk bounded by free-slip surface at the top. For the fine-block, we used $\Delta^{+}_{fine}=3.95$
in wall units. Due to the use of link-bounce back method for implementation of wall boundary
condition, the first lattice node is located at a distance of $\Delta^{+}_{nw}=\Delta^{+}/2$,
which in our case is $1.98$. Hence, the computations are expected to fairly resolve the small-scale
turbulent structures. For the coarse-block, we used $\Delta^{+}_{coarse}=2\times \Delta^{+}_{fine}=7.90$.
The computational domain is discretized by $256\times 128\times 12$ grid nodes and $128\times 64\times 19$
grid nodes in the fine and coarse blocks, respectively.

The initial mean velocity is specified to satisfy the $1/7^{th}$
power law~\cite{pope00}, while initial perturbations satisfying
divergence free velocity field~\cite{lam92}. The density field is
taken to be $\rho=\rho_0=1.0$ for the entire domain. The precise
form of the initial fields may not affect the turbulence
statistics, but can have significant influence on the convergence of
the solution to statistically steady state. In particular, the above
choice of initial fields would enable a rapid convergence to the
statistically steady state solution of the fluctuating fields
obtained by GLBE with forcing term. We employed the consistent initialization
procedure for the distribution functions or moments~\cite{mei06}, that properly initializes
non-equilibrium moments of the GLBE in the presence of non-uniform hydrodynamic fields, such as those
mentioned above.

\subsubsection{Computational Procedure}
Using $\overrightarrow{F}=-\frac{dp}{dx}\widehat{x}=\frac{\tau_w}{H}\widehat{x}=\frac{\rho
u_{*}^2}{H}\widehat{x}$ as the driving force, LES using the GLBE is performed. The collision step, including
the forcing term, is computed in moment space, while the streaming step is carried out in its natural
velocity space, and the result of these two steps provides the grid-filtered hydrodynamic fields.
The grid-filtered strain rate tensor $\overline{S}_{ij}$ is obtained locally from non-equilibrium moments as shown in
Appendix~\ref{app:momentcomponents}, which was also found to provide improved numerical stability on
coarser grids when used in lieu of finite-differencing of the velocity fields. The relaxation process in the collision step is carried out by computing the ``effective" relaxation times for hydrodynamic moments obtained from a variable coefficient Smagorinsky SGS eddy-viscosity model, whose values are determined by a dynamic procedure. The steps to accomplish this is briefly summarized as follows.

Based on these grid-filtered quantities and considering $\widetilde{\overline{\Delta}}/\overline{\Delta}=2$ where $\overline{\Delta}=\delta_x$, discrete trapezoidal test-filters, as discussed in Sec.~\ref{subsec:filters}, are applied to obtain the following quantities: $\widetilde{\overline{u}}_i$, $\widetilde{\overline{u}_i\overline{u}_j}$, $\widetilde{\overline{S}}_{ij}$, $|\widetilde{\overline{S}}|$ and
$\widetilde{|\overline{S}|\overline{S}_{ij}}$. Once they are known, tensors such as $L_{ij}$ and $M_{ij}$ used in the
dynamic procedure (Eqs.~(\ref{eq:resolvedstress} and~\ref{eq:MStressexpression}), respectively, see Sec.~\ref{sec:dynamicSGS}) are computed. The model coefficient $C$ is, then, obtained by averaging the contracted tensor products, as shown in Eq.~(\ref{eq:Ccoefficient}), along the homogeneous $x$ and $y$ directions, i.e. horizontal planes, to maintain numerical stability. Moreover, near free-slip surfaces, we specify that there is no change in the model
coefficient around the top surface to ensure stability. Other SGS models, such as the DTM whose implementation details in GLBE framework are provided in the Appendix~\ref{app:dynamicmixedmodels}, could be used to relax this assumption, which is a subject for a future work. Finally, the SGS eddy-viscosity is obtained from Eq.~(\ref{eq:eddyviscosity}) to provide the ``effective"
relaxation times for the collision step. The GLBE computations are carried out until stationary turbulence statistics are obtained, as measured by the invariant Reynolds stresses profiles. This initial run was carried out for a duration of $50T^{*}$, where $T^{*}=H/u_{*}$ is the characteristic time scale. The averaging of various flow quantities was carried out in time as well as in space in the homogeneous directions by an additional run for a period of $12T^{*}$.

\subsubsection{Results}
\begin{figure}
\includegraphics[width = 105mm]{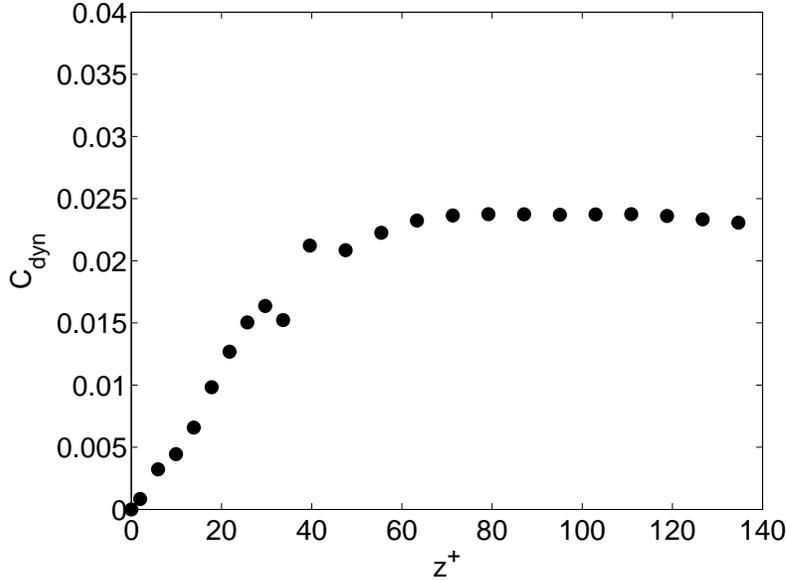}
\caption{\label{fig:Cdyn180cut5} Computed variation of model coefficient for fully-developed turbulent channel flow with a top slip surface at a shear Reynolds number of $Re_{*}=180$ using a dynamic procedure with multiblock GLBE.}
\end{figure}

Figure~\ref{fig:Cdyn180cut5} shows the variation of the SGS model coefficient $C_{dyn}=C$ along the wall normal direction, where the distance is scaled in the wall coordinates $Z^{+}$, where $Z^{+}=Z/\delta_{\nu}$ and where $\delta_{\nu}$ is the viscous length scale defined earlier.
The model coefficient is seen to be nearly a constant in the bulk or outer flow region, where turbulence tends towards becoming statistically more homogeneous and isotropic. In particular, the value of $C_{dyn}$ in these regions is found to be about $0.024$. This is excellent agreement with Germano \emph{et al}. (1991)~\cite{germano91} based on the solution of filtered Navier-Stokes equations, who found a value of $0.023$. The eddy-viscosity given in Eq.~(\ref{eq:eddyviscosity}), is also written in terms of the Smagorinsky ``constant" $C_S$ as $\nu_t=C_S^2\overline{\Delta}^2|\overline{S}|$. Thus, $C_S=\sqrt{C}$ and we find $C_S=0.155$ in the bulk region, which is within the range of $0.10-0.16$ reported for various classes of flows~\cite{galperin93}. Near $Z^+=40$, the grids transition from coarse to fine block, where some oscillations in the values of $C$ are noticed, which could be removed by an additional averaging. Closer to the wall ($Z^{+}<30$), in the inner-layer region, the model coefficient is found to decrease monotonically with decreasing distance from the wall $Z^{+}$ and approaching zero very close to the wall. This trend is consistent with near-wall turbulence physics, according to which turbulence is statistically anisotropic and its scales become smaller towards the wall. Thus, the GLBE using dynamic Smagorinsky SGS model is not only able to self-consistently predict the bulk value of the model coefficient automatically and accurately, but also its variation near the wall without resorting to any empirical approach such as the van Driest damping function~\cite{vanDriest56}. Such predictive capabilities are among the major assets of dynamic SGS modeling using GLBE.

\begin{figure}
\includegraphics[width = 105mm]{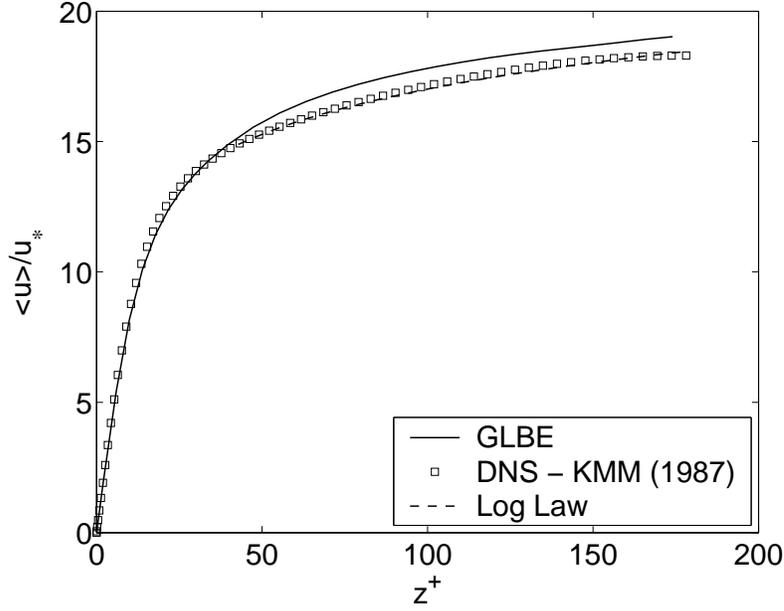}
\caption{\label{fig:ubar180a} Mean velocity in fully-developed turbulent channel flow with a top slip surface at a shear Reynolds number of $Re_{*}=180$ using a dynamic procedure with multiblock GLBE. Comparison with DNS of Kim, Moin and Moser (1987).}
\end{figure}

Figure~\ref{fig:ubar180a} shows the computed mean velocity profile,
normalized by the shear velocity $u_{*}$, as a function of the
distance from the wall given in wall units, i.e. $Z^{+}$.
Also plotted are the DNS data by Kim, Moin and Moser (1987) based on a
spectral method~\cite{kim87}. The computed velocity
profile follows fairly closely with DNS data, with about $5\%$
difference. Such differences are characteristic of LES, which employ
relatively coarser grids than DNS, and which also generally depends
on the numerical dissipation of the computational approach for LES
(see e.g., Ref.~\cite{choi00,gullbrand03}).

\begin{figure}
\includegraphics[width = 105mm]{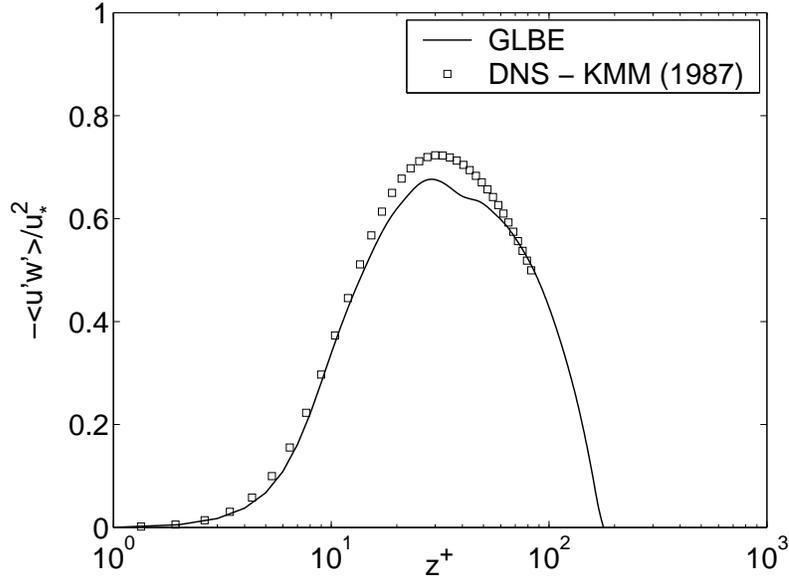}
\caption{\label{fig:rs180} Mean Reynolds stress in fully-developed turbulent channel flow with a top slip surface at a shear Reynolds number of $Re_{*}=180$ using a dynamic procedure with multiblock GLBE. Comparison with DNS of Kim, Moin and Moser (1987).}
\end{figure}

The Reynolds stress, normalized by the wall-shear stress, is
presented in Fig.~\ref{fig:rs180} in semi-log scale and
compared with the DNS data of Ref.~\cite{kim87} obtained from the
direct solution of incompressible Navier-Stokes equations (NSE).
It is found that, in general, the computed Reynolds stress is in reasonably good agreement with
DNS. The computed values are somewhat under predicted near $Z^{+}=40$, where the transition between
coarse to fine blocks of grid nodes occur. It is well known that where different types of grids interface,
certain amount of numerical effects are inevitable, as was found in certain flow problems~\cite{rohde06}.
Nevertheless, it is clear that the computations are able to predict the Reynolds stress variation both qualitatively
and quantitatively in a reasonably accurate manner.

\begin{figure}
\includegraphics[width = 105mm]{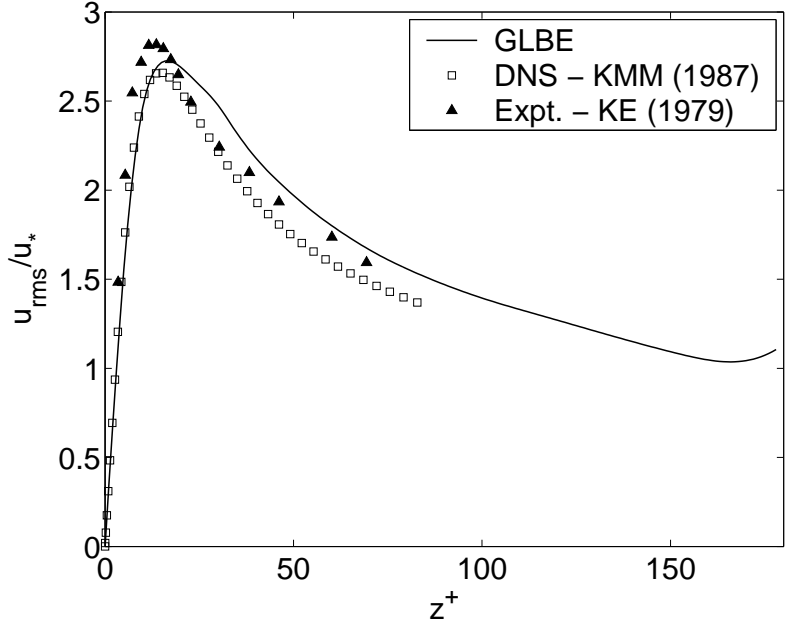}
\caption{\label{fig:urms180} Root-mean-square(rms) streamwise velocity fluctuations in fully-developed turbulent channel flow with a top slip surface at a shear Reynolds number of $Re_{*}=180$ using a dynamic procedure with multiblock GLBE. Comparison with DNS of Kim, Moin and Moser (1987) and measurements of Kreplin and Eckelmann (1979).}
\end{figure}

\begin{figure}
\includegraphics[width = 105mm]{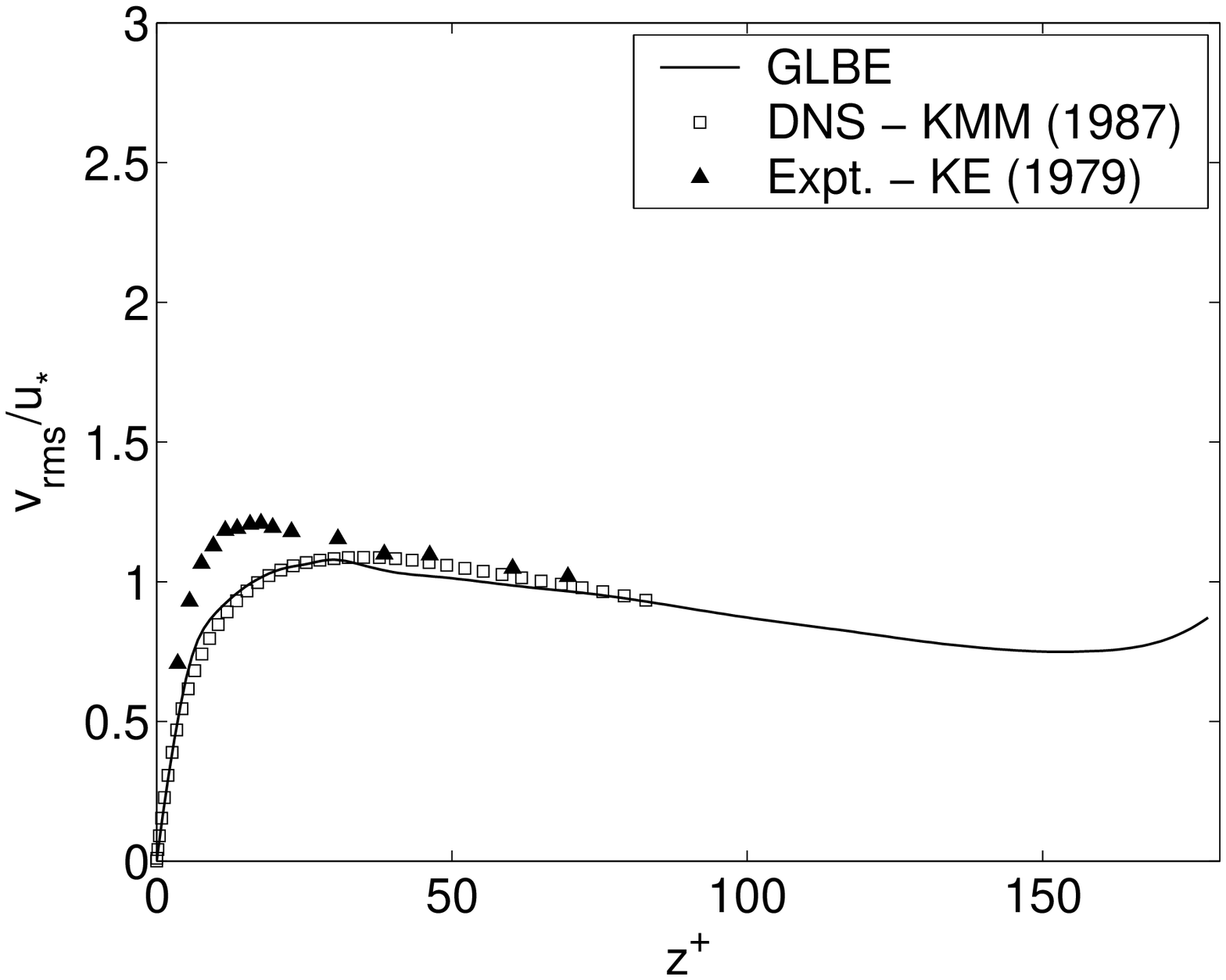}
\caption{\label{fig:vrms180} Root-mean-square(rms) spanwise velocity fluctuations in fully-developed turbulent channel flow with a top slip surface at a shear Reynolds number of $Re_{*}=180$ using a dynamic procedure with multiblock GLBE. Comparison with DNS of Kim, Moin and Moser (1987) and measurements of Kreplin and Eckelmann (1979).}
\end{figure}

\begin{figure}
\includegraphics[width = 105mm]{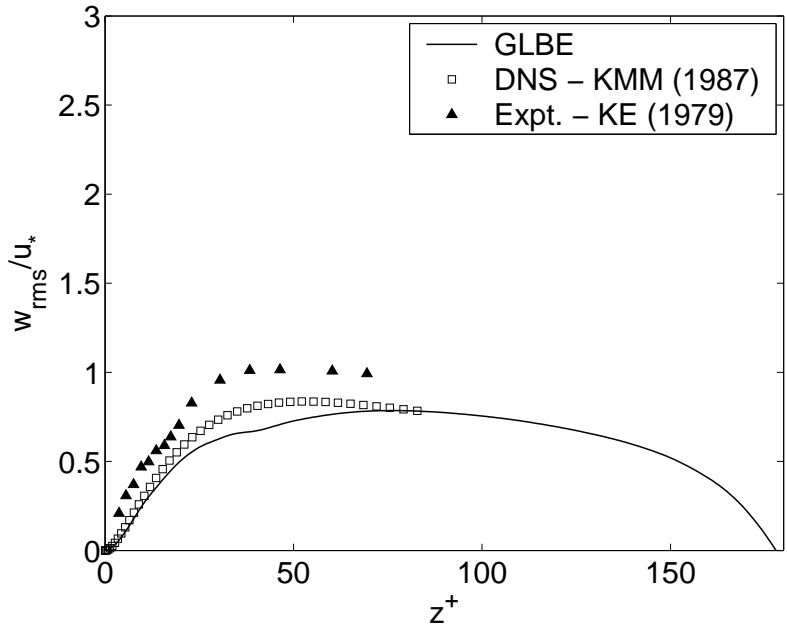}
\caption{\label{fig:wrms180} Root-mean-square(rms) wall normal velocity fluctuations in fully-developed turbulent channel flow with a top slip surface at a shear Reynolds number of $Re_{*}=180$ using a dynamic procedure with multiblock GLBE. Comparison with DNS of Kim, Moin and Moser (1987) and measurements of Kreplin and Eckelmann (1979).}
\end{figure}

Second-order statistics, such as turbulence intensities are important measures of turbulent activity and let us
now assess how the LES computations using the GLBE with a dynamic SGS model are able to predict such quantities
in the near-wall region. Figures~\ref{fig:urms180},
\ref{fig:vrms180}
and \ref{fig:wrms180}
show comparisons of the components of the root-mean-square (rms)
streamwise, spanwise and wall-normal velocity fluctuations,
respectively, computed using the GLBE with data from DNS based on the
NSE~\cite{kim87} and experimental measurements~\cite{kreplin79}. Again, the computed results
are in good agreement with prior data. It is interesting to note among the three components of
turbulence intensities, only the wall normal component is the most sensitive to the numerical effects
due to coarse-to-fine grid interfaces, which are in any case reasonably small.

\subsection{Turbulent Channel Flow at $Re_{*}=395$}
It is important to know whether the approach discussed in this paper is able to scale well and reproduce
turbulence statistics at higher Reynolds numbers. In this regard, we assess the LES approach based on
GLBE using dynamic SGS modeling at a higher shear Reynolds number of $395$, for which DNS data is available
from a more recent work by Moser, Kim and Mansour (1999)~\cite{moser99}. To the best of authors' knowledge, LES using LBM, even
with constant Smagorinsky SGS model, at such a higher Reynolds number and their comparison with the DNS has not previously been reported in the literature.
In order to properly sample sufficient number of wall-layer streaks and exclude end effects of
two-point velocity correlations, we choose a larger computational domain than in the previous case,
with a size of $5.8H\times 2.9H\times H$. The computational domain is discretized in terms of four blocks,
with the finest grid block near walls, with a spacing of $\Delta^{+}=3.95$. The first grid node is located
at $\Delta_{nw}^{+}=\Delta^{+}/2=1.96$ due to the use of half-way link bounce back, and hence can resolve the
smaller near-wall turbulence structures properly. The four blocks are resolved by means of the following number of
grid nodes: $576\times 288 \times 12$, $288\times 144 \times 12$, $144\times 72 \times 15$ and $288\times 144 \times 8$, and
the GLBE computations are carried out till stationary turbulence are observed, followed by an additional run for a duration of
$13H/u_{*}$ for obtaining turbulence statistics.

\begin{figure}
\includegraphics[width = 105mm]{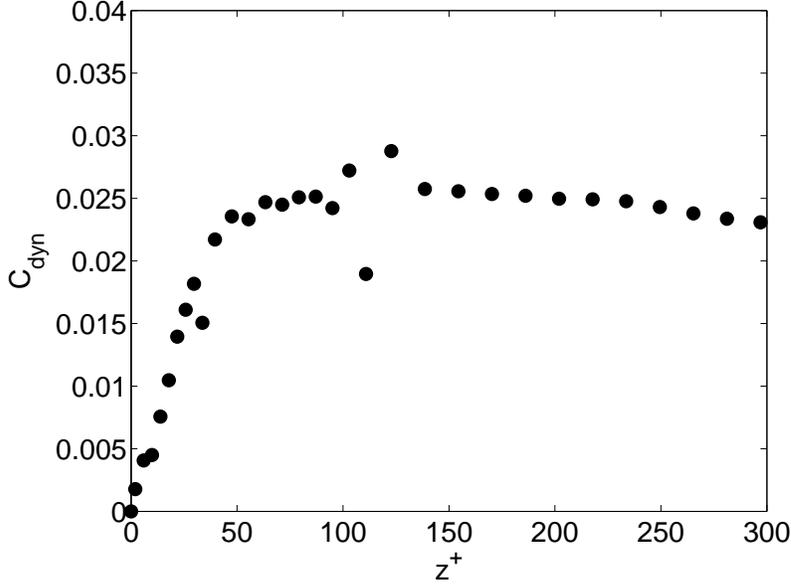}
\caption{\label{fig:Cdyn395cut4} Computed variation of model coefficient for fully-developed turbulent channel flow with a top slip surface at a shear Reynolds number of $Re_{*}=395$ using a dynamic procedure with multiblock GLBE.}
\end{figure}

Figure~\ref{fig:Cdyn395cut4} shows the computed variation of the model coefficient $C_{dyn}=C$ as a function of the distance
from the wall $Z^{+}$ for $Re_{*}=395$.
Again, the dynamic LES approach is able to predict the value of the model coefficient in the bulk or outer flow region
accurately. Moreover, it is also able to reproduce the near-wall trend of $C_{dyn}$ well, viz., its reduction as the distance
from the wall is decreased. It is also noticed that around regions, where there is grid block transition, some oscillations in
$C_{dyn}$ are present. The magnitude of these oscillations appear to progressively increases with grid transitions involving coarser grid blocks, with the maximum occurring at around $Z^{+}=120$. However, they are generally small and do not seem to affect the turbulent statistics as shown below.

\begin{figure}
\includegraphics[width = 105mm]{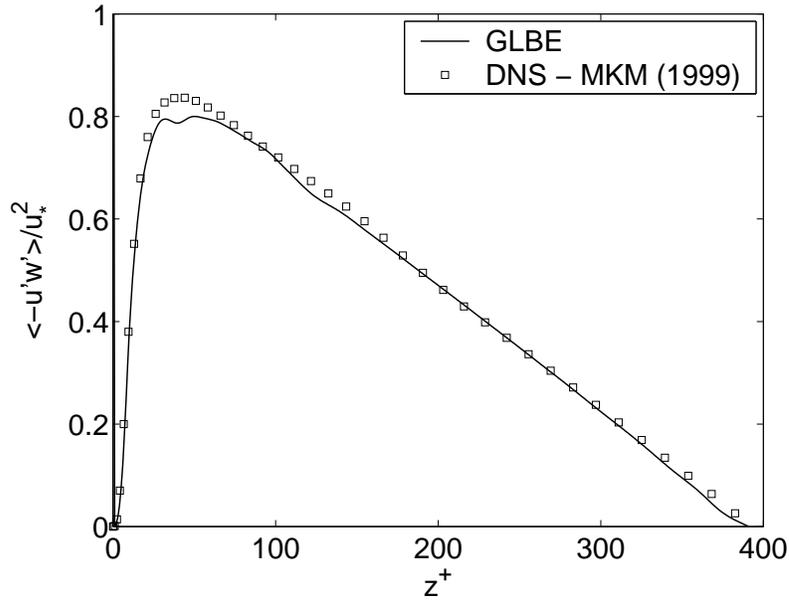}
\caption{\label{fig:rs395} Mean Reynolds stress in fully-developed turbulent channel flow with a top slip surface at a shear Reynolds number of $Re_{*}=395$ using a dynamic procedure with multiblock GLBE. Comparison with DNS of Moser, Kim and Mansour (1999).}
\end{figure}

Figure~\ref{fig:ubar395} shows the computed mean velocity profile, normalized by the shear velocity $u_{*}$,
\begin{figure}
\includegraphics[width = 105mm]{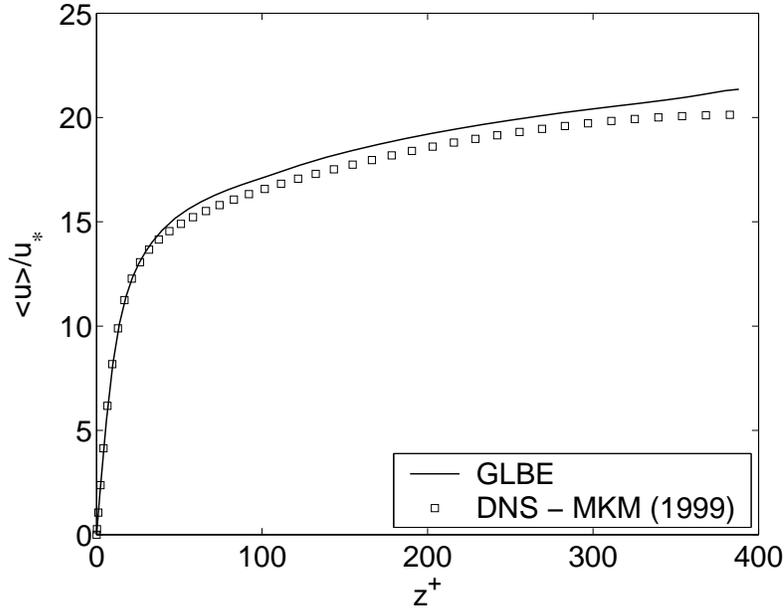}
\caption{\label{fig:ubar395} Mean velocity in fully-developed turbulent channel flow with a top slip surface at a shear Reynolds number of $Re_{*}=180$ using a dynamic procedure with multiblock GLBE. Comparison with DNS of Moser, Kim and Mansour (1999).}
\end{figure}
while the Reynolds stress, normalized by the wall shear stress, for $Re_{*}=395$ is presented in Fig.~\ref{fig:rs395}.
The DNS data of Moser, Kim and Mansour (1999)~\cite{moser99} are also plotted in these figures for comparison. Notice that the peak Reynolds stress values are higher for the $Re_{*}=395$ case as compared to the $Re_{*}=180$, reflecting greater turbulent activity at the higher Reynolds number. It is found that the GLBE LES approach using dynamic SGS model is able to reproduce these quantities reasonably well.

\begin{figure}
\includegraphics[width = 105mm]{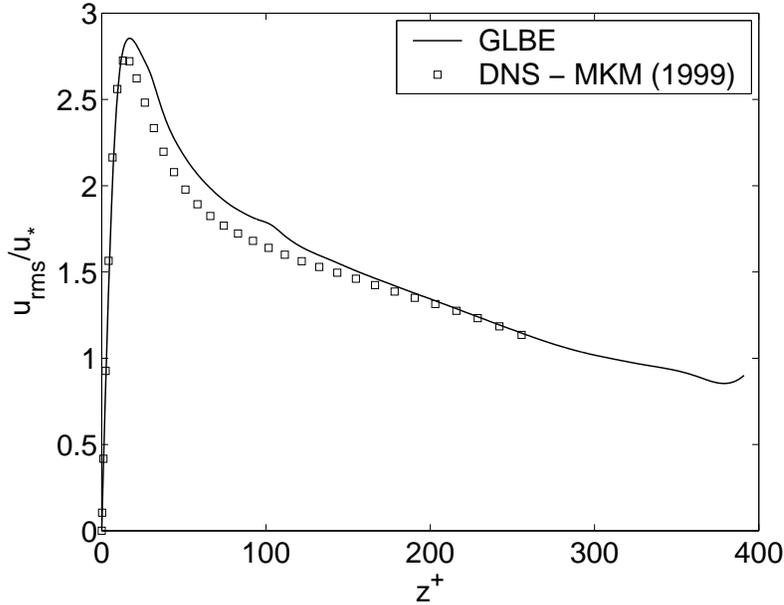}
\caption{\label{fig:urms395} Root-mean-square(rms) streamwise velocity fluctuations in fully-developed turbulent channel flow with a top slip surface at a shear Reynolds number of $Re_{*}=395$ using a dynamic procedure with multiblock GLBE. Comparison with DNS of Moser, Kim and Mansour (1999).}
\end{figure}

\begin{figure}
\includegraphics[width = 105mm]{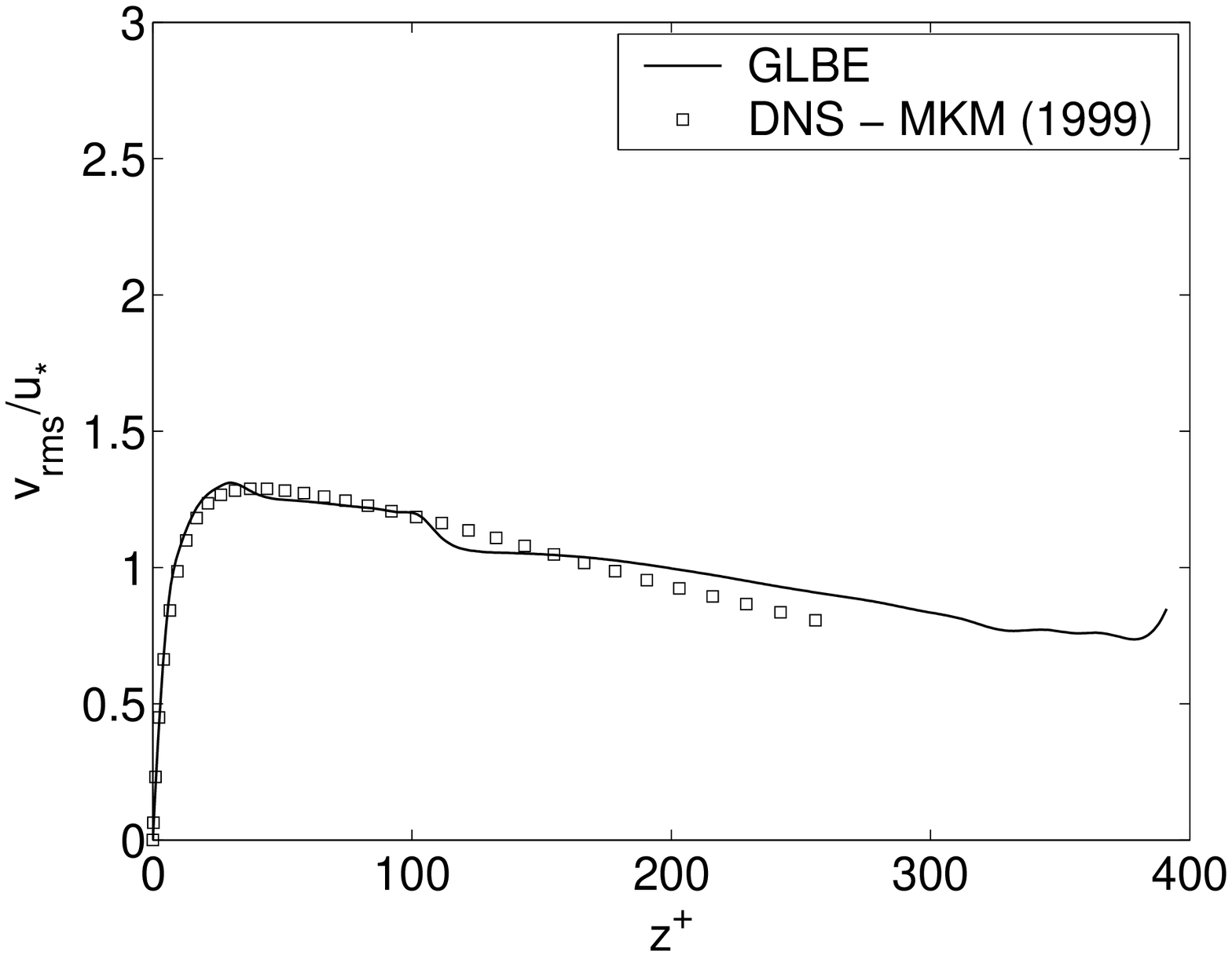}
\caption{\label{fig:vrms395} Root-mean-square(rms) spanwise velocity fluctuations in fully-developed turbulent channel flow with a top slip surface at a shear Reynolds number of $Re_{*}=395$ using a dynamic procedure with multiblock GLBE. Comparison with DNS of Moser, Kim and Mansour (1999).}
\end{figure}

\begin{figure}
\includegraphics[width = 105mm]{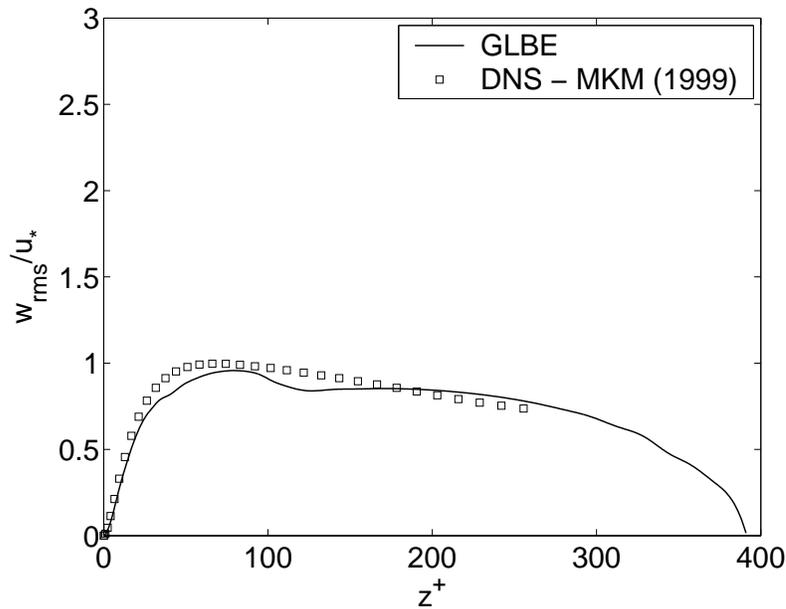}
\caption{\label{fig:wrms395} Root-mean-square(rms) wall normal velocity fluctuations in fully-developed turbulent channel flow with a top slip surface at a shear Reynolds number of $Re_{*}=395$ using a dynamic procedure with multiblock GLBE. Comparison with DNS of Moser, Kim and Mansour (1999).}
\end{figure}

Let us now present the root-mean-square (rms) velocity fluctuations in the streamwise, spanwise and wall normal directions in
Figs.~\ref{fig:urms395},
\ref{fig:vrms395} and
\ref{fig:wrms395}, respectively.
These computed quantities are normalized by the shear velocity and compared with the DNS data~\cite{moser99}. It is noticed that the computed peak values of all the components of turbulent intensities are higher for $Re_{*}=395$ than for $Re_{*}=180$. Moreover, the computed profiles are in good agreement with DNS.

\begin{figure}
\includegraphics[width = 105mm]{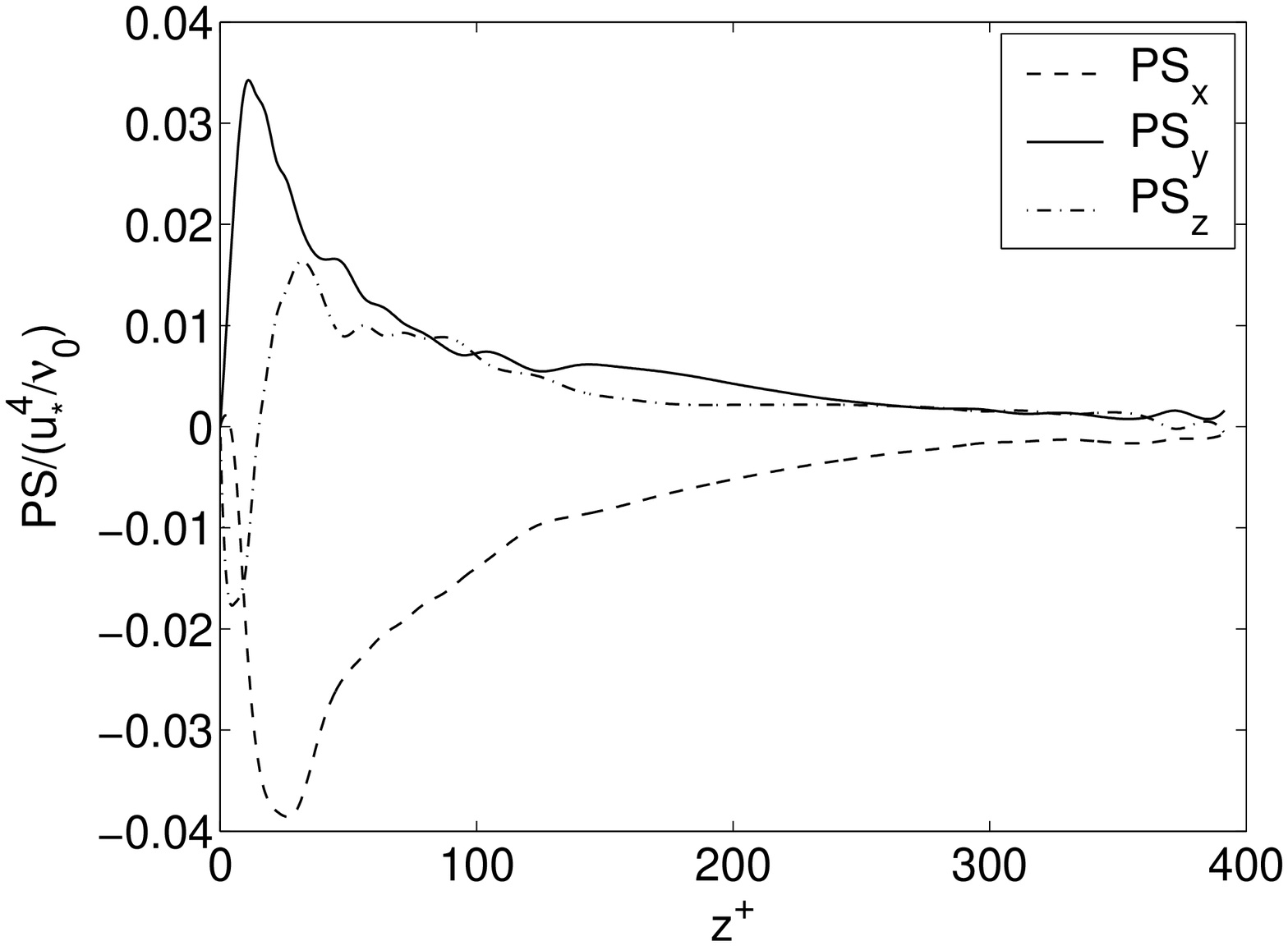}
\caption{\label{fig:ps395} Components of pressure strain correlations in fully-developed turbulent channel flow with a top slip surface at a shear Reynolds number of $Re_{*}=395$ using a dynamic procedure with multiblock GLBE.}
\end{figure}

An important indication of how turbulence kinetic energy is transferred
among the components is provided by the pressure-strain (PS) correlations.
Their components are: $PS_x=\langle p^{'}\partial_x u_x^{'}\rangle$,
$PS_y=\langle p^{'}\partial_y u_y^{'}\rangle$, and $PS_z=\langle
p^{'}\partial_z u_z^{'}\rangle$, where the prime denotes
fluctuations and the brackets refer to averaging (along homogeneous
spatial directions and time). Figure~\ref{fig:ps395} shows
the components of the computed PS correlations.
They exhibit the expected behavior close to the wall, including the
transfer of energy from the wall-normal component to the other two
components near the wall - a phenomenon termed as splatting or
impingement~\cite{moin82}.

\begin{figure}
\includegraphics[width = 105mm]{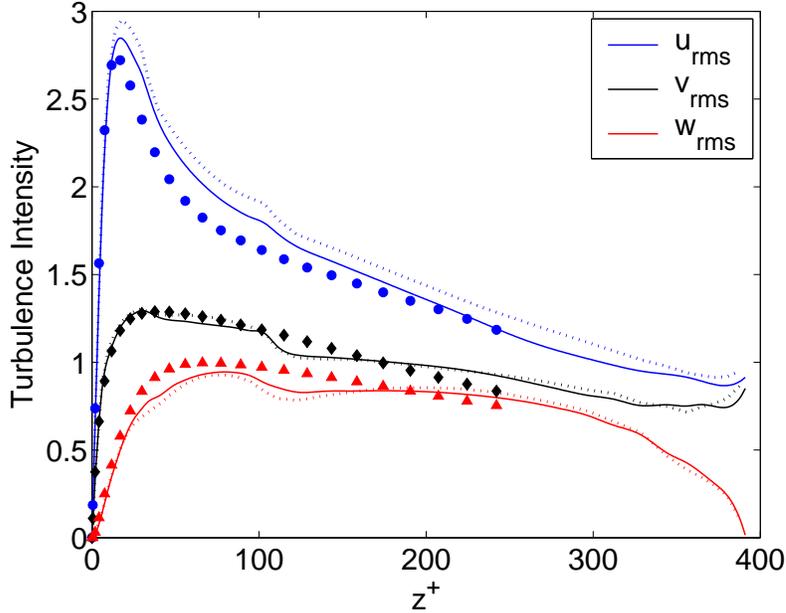}
\caption{\label{fig:smag-const-dyn3} Comparison between constant (dashed) and dynamic (solid) Smagorinsky models implemented in a multiblock GLBE framework for LES of fully-developed turbulent channel flow with a top slip surface at a shear Reynolds number of $Re_{*}=395$. Symbols are DNS data of Moser, Kim and Mansour (1999). For the constant Smagorinksy model, $C_s=0.15$ with van Driest damping using a coefficient of $A^{+}=26$ were considered.}
\end{figure}

It will be interesting to compare the computed results using the dynamic Smagorinsky SGS model
with those using the ``constant" Smagorinsky SGS model with an \emph{ad hoc} van Driest damping function~\cite{vanDriest56}.
As an illustration, Fig.~\ref{fig:smag-const-dyn3} shows a comparison of the components of
turbulence intensities with both these SGS models. Also, plotted in symbols is the DNS data~\cite{moser99}.
In the case of ``constant" Smagorinsky SGS model, we consider $C_{S}=0.15$ with the following damping function:
$\Delta=\delta_x\left[1-\exp\left( -\frac{z^{+}}{A^{+}}\right)\right]$, where $A^{+}=26$. We find that even with the commonly used values of empirical constants, the results with the dynamical SGS model is in better agreement with the DNS. It may be noted that without the use of damping function, the ``constant" Smagorinsky SGS model would have given grossly inaccurate results for this problem. In any case, the important point is that the dynamic SGS model, when used with the GLBE for LES, does not require any \emph{a priori} information regarding the values or the behavior of the model coefficient, which are computed from the local behavior of the turbulent fields itself.

\section{\label{sec:summary}Summary and Conclusions}
In this paper we discussed the incorporation of a dynamic procedure for large eddy simulation (LES)
of complex turbulent flows using the generalized lattice Boltzmann equation (GLBE) with forcing
term based on multiple relaxation times. The use of dynamic procedure, which samples super-grid or ``test"
scale turbulent dynamics and assuming scale-invariance between two levels to compute the model coefficient,
largely eliminates empiricism required to close the SGS turbulence models, particularly for
inhomogeneous and anisotropic turbulent flows. Eddy-viscosity type dynamic Smagorinsky model (DSM) is
employed as the SGS model in this work. The solution of the GLBE, which has significantly improved fidelity
and numerical stability when compared to the SRT-LBE due to its ability to separate relaxation time scales
of hydrodynamic and kinetic modes, yields the grid-filtered hydrodynamic fields, when ``effective" relaxation times
based on the dynamic SGS model is employed. Variable resolutions, to accommodate different turbulent length scales
with the finest near the walls, are considered by the use of conservative, locally refined multiblock gridding approach. The
grid-filtered strain-rate tensor used in the SGS model is computed locally from the non-equilibrium
moments of the GLBE, which has superior stability characteristics when compared to the use of simple
finite-differencing of the velocity field on coarser grids. To allow efficient implementation, discrete trapezoidal
filters are employed to obtain the test-filtered quantities in the dynamic procedure.

Numerical simulation results using the DSM are obtained for the canonical fully-developed turbulent channel flow at shear Reynolds numbers $Re_{*}$ of $180$ and $395$. It is demonstrated that the use of the multiblock GLBE in conjunction with dynamic procedure for the SGS model is able to automatically compute the values and variation of the model coefficient $C$. In particular, without the use of any \emph{ad hoc} approach, the computations reproduce the value of $C$ equal to $0.024$ (or $C_{S}=0.155)$ in the bulk flow region, which reduces and approaches to zero towards to the wall, reflecting near-wall turbulence physics. The computed turbulent statistics such as the root-mean-square velocity fluctuations and Reynolds stresses for both cases of Reynolds numbers are in good agreement with prior direct numerical simulations (DNS) and experimental data, as applicable. In particular, the peaks of the second-order statistics scale to higher values at higher $Re_{*}$, which the
computations are found to reproduce quantitatively well. While the focus of this paper, as a first step, is on the use of dynamic procedure for eddy-viscosity based models in the LBM framework, it also outlines the details of implementation of more general scale-similarity based dynamic SGS models. It appears that the use of dynamic procedure in the LBM, particularly with the GLBE, is a promising approach for LES of complex turbulent flows.

\section*{\label{acknowledgements}Acknowledgements}
This work was performed under the auspices of the National
Aeronautics and Space Administration (NASA) under Contract
No.~NNL07AA04C. Computational resources were provided by the
National Center for Supercomputing Applications (NCSA) under
Award CTS 060027 and the Office of Science of DOE under Contract
DE-AC03-76SF00098.

\newpage

\appendix

\section{\label{app:momentcomponents} Moments, Equilibrium Moments, Moment-projections of Source Terms for D3Q19 Model}
The components of the various elements in the moments are as
follows~\cite{dhumieres02}:

$\widehat{f}_0 = \rho, \widehat{f}_1= e,
\widehat{f}_2 = e^2, \widehat{f}_3 = j_x,\widehat{f}_4 =
q_x,\widehat{f}_5 = j_y, \widehat{f}_6 = q_y, \widehat{f}_7 = j_z,
\widehat{f}_8 = q_z, \widehat{f}_9 = 3p_{xx},\widehat{f}_{10} =
3\pi_{xx},\widehat{f}_{11} = p_{ww},\widehat{f}_{12} =
\pi_{ww},\widehat{f}_{13} = p_{xy},\widehat{f}_{14} =
p_{yz},\widehat{f}_{15} = p_{xz},\widehat{f}_{16} =
m_x,\widehat{f}_{17} = m_y,\widehat{f}_{18} = m_z$. Here, $\rho$ is
the density, $e$ and $e^2$ represent kinetic energy that is
independent of density and square of energy, respectively; $j_x$,
$j_y$ and $j_z$ are the components of the momentum, i.e. $j_x = \rho
u_x$, $j_y = \rho u_y$, $j_z = \rho u_z$, $q_x$, $q_y$, $q_z$ are
the components of the energy flux, and $p_{xx}$, $p_{xy}$, $p_{yz}$
and $p_{xz}$ are the components of the symmetric traceless viscous
stress tensor; The other two normal components of the viscous stress
tensor, $p_{yy}$ and $p_{zz}$, can be constructed from $p_{xx}$ and
$p_{ww}$, where $p_{ww} = p_{yy} - p_{zz}$. Other moments include
$\pi_{xx}$, $\pi_{ww}$, $m_x$, $m_y$ and $m_z$. The first two of
these moments have the same symmetry as the diagonal part of the
traceless viscous tensor $p_{ij}$, while the last three vectors are
parts of a third rank tensor, with the symmetry of $j_kp_{mn}$.

The corresponding components of the equilibrium moments, which are
functions of conserved moments, i.e. density $\rho$ and momentum
$\overrightarrow{j}$, are as follows~\cite{dhumieres02}:

$\widehat{f}_0^{eq} = \rho, \widehat{f}_1^{eq} \equiv
e^{eq}=-11\rho+19\frac{\overrightarrow{j}\cdot\overrightarrow{j}}{\rho},
\widehat{f}_2^{eq} \equiv
e^{2,eq}=3\rho-\frac{11}{2}\frac{\overrightarrow{j}\cdot\overrightarrow{j}}{\rho},
\widehat{f}_3^{eq} = j_x,\widehat{f}_4^{eq} \equiv
q_x^{eq}=-\frac{2}{3}j_x,\widehat{f}_5^{eq} = j_y,
\widehat{f}_6^{eq} \equiv q_y^{eq}=-\frac{2}{3}j_y,
\widehat{f}_7^{eq} = j_z, \widehat{f}_8^{eq} \equiv
q_z^{eq}=-\frac{2}{3}j_z, \widehat{f}_9^{eq} \equiv
3p_{xx}^{eq}=\frac{\left[3j_x^2-\overrightarrow{j}\cdot\overrightarrow{j}
\right]}{\rho},\widehat{f}_{10}^{eq} \equiv
3\pi_{xx}^{eq}=3\left(-\frac{1}{2}p_{xx}^{eq}
\right),\widehat{f}_{11}^{eq} \equiv
p_{ww}^{eq}=\frac{\left[j_y^2-j_z^2
\right]}{\rho},\widehat{f}_{12}^{eq} \equiv
\pi_{ww}^{eq}=-\frac{1}{2}p_{ww}^{eq},\widehat{f}_{13}^{eq} \equiv
p_{xy}^{eq}=\frac{j_xj_y}{\rho},\widehat{f}_{14}^{eq} \equiv
p_{yz}^{eq}=\frac{j_yj_z}{\rho},\widehat{f}_{15}^{eq} \equiv
p_{xz}^{eq}=\frac{j_xj_z}{\rho},\widehat{f}_{16}^{eq} =
0,\widehat{f}_{17}^{eq} = 0,\widehat{f}_{18}^{eq} = 0$.

The components of the source terms in moment space are functions of
external force $\overrightarrow{F}$ and velocity fields
$\overrightarrow{u}$, respectively, as follows~\cite{premnath08}:

$\widehat{S}_0 = 0, \widehat{S}_1 = 38(F_xu_x+F_yu_y+F_zu_z),
\widehat{S}_2 = -11(F_xu_x+F_yu_y+F_zu_z), \widehat{S}_3= F_x,
\widehat{S}_4 =-\frac{2}{3}F_x, \widehat{S}_5=F_y, \widehat{S}_6
=-\frac{2}{3}F_y, \widehat{S}_7=F_z, \widehat{S}_8 =
-\frac{2}{3}F_z, \widehat{S}_9 = 2(2F_xu_x-F_yu_y-F_zu_z),
\widehat{S}_{10} = -(2F_xu_x-F_yu_y-F_zu_z), \widehat{S}_{11} =
2(F_yu_y-F_zu_z), \widehat{S}_{12} = -(F_yu_y-F_zu_z),
\widehat{S}_{13} = (F_xu_y+F_yu_x), \widehat{S}_{14} = (F_yu_z+F_zu_y),
\widehat{S}_{15} = (F_xu_z+F_zu_x), \widehat{S}_{16} = 0,
\widehat{S}_{17} = 0, \widehat{S}_{18} = 0$.

The components of the strain rate tensor used in subgrid scale (SGS)
turbulence model at the grid-filter level can be written explicitly in terms of
non-equilibrium moments augmented by moment-projections of source
terms as~\cite{premnath08}
\begin{eqnarray}
S_{xx}&\approx&-\frac{1}{38\rho} \left[
s_{1}\widehat{h}_{1}^{(neq)}+19s_{9}\widehat{h}_{9}^{(neq)}  \right], \label{eq:sxx}\\
S_{yy}&\approx&-\frac{1}{76\rho} \left[
2s_{1}\widehat{h}_{1}^{(neq)}-19\left(s_{9}\widehat{h}_{9}^{(neq)}-3s_{11}\widehat{h}_{11}^{(neq)}\right)
\right],\label{eq:syy}\\
S_{zz}&\approx&-\frac{1}{76\rho} \left[
2s_{1}\widehat{h}_{1}^{(neq)}-19\left(s_{9}\widehat{h}_{9}^{(neq)}+3s_{11}\widehat{h}_{11}^{(neq)}\right)
\right],\label{eq:szz}\\
S_{xy}&\approx&-\frac{3}{2\rho}
s_{13}\widehat{h}_{13}^{(neq)}, \label{eq:sxy}\\
S_{yz}&\approx&-\frac{3}{2\rho}
s_{14}\widehat{h}_{14}^{(neq)}, \label{eq:syz}\\
S_{xz}&\approx&-\frac{3}{2\rho}
s_{15}\widehat{h}_{15}^{(neq)},\label{eq:sxz}
\end{eqnarray}
where
\begin{equation}
\widehat{h}_{\alpha}^{(neq)}=\widehat{f}_{\alpha}-\widehat{f}_{\alpha}^{eq}+\frac{1}{2}\widehat{S}_{\alpha},\qquad
\alpha \in \{1,9,11,13,14,15\}
\end{equation}
Here, $\widehat{f}_{\alpha}$, $\widehat{f}^{eq}_{\alpha}$ and
$\widehat{S}_{\alpha}$ are components of moments, their local
equilibria, and moment-projections of source terms due to external
forces, respectively, which are given above. $s_{\alpha}$ are
elements of the collision matrix
$\widehat{\Lambda}=diag(s_0,s_1,\ldots,s_{18})$ in moment space. The
expressions for the strain rate tensor are generalizations of those
given in Yu \emph{et al}.~\cite{yu06}

\section{\label{app:dynamicmixedmodels} Incorporation of Dynamic Mixed Model (DMM) and Dynamic Two Parameter (DTM) Subgrid Scale Models in Lattice-Boltzmann Method}
It is well recognized that improvements to the dynamic Smagorinsky model (DSM) is
required to properly account for various aspects of subgrid scale turbulence physics
in general situations. For example, while DSM is generally adequate for performing large-eddy
simulation (LES) of bulk as well as near-wall turbulent flows, it is less satisfactory near the free-slip
or free-surfaces. This is presumably because of the assumption in the DSM that turbulence
dissipation is in equilibrium with production, leading to the eddy-viscosity type models, which
implies the alignment of the principal axes of the SGS stress tensor with those of the resolved
strain rate tensor. Moreover, when the dynamic coefficient $C$ is computed locally, it exhibits
large fluctuations leading to excessive energy backscatter and numerical instability. Thus, the
DSM can accurately predict mean flow quantities when an averaged value of the coefficient is
employed, but is inadequate for representation of the local quantities.

Thus, Zang \emph{et al}. (1993)~\cite{zang93} proposed to circumvent these problems by employing a mixed
model, which also contains contributions from the scale-similarity based stresses or the
modified Leonard stress terms (as originally proposed by Bardina \emph{et al}. (1980)~\cite{bardina80})
and do not assume the alignment of the SGS stress and the resolved
strain rate tensors. The proportionality constant for the Leonard stress terms is set to 1.0 in this
model. By alleviating the burden on the variation of the coefficient $C$ to represent turbulence
physics, this dynamic mixed model (DMM) provided improved results, with significantly less
fluctuations in the model coefficient. A further improvement, particularly in the context of LES
of free-surface turbulence, was provided by the dynamic two-parameter model (DTM) of Salvetti
and Banerjee (1995)~\cite{salvetti95}. In this model, an additional parameter $K$ is introduced as a proportionality
constant to the modified Leonard stress term in the mixed based model. Both the
coefficients $C$ and $K$ can then be obtained locally by means of a least-squares technique, as
formulated by Lilly (1992)~\cite{lilly92}. We now discuss the development of a procedure for incorporation of
DTM in the framework of lattice Boltzmann method (LBM).

The anisotropic part of the SGS stress in the DTM can be represented as
\begin{equation}
\tau_{ij}-\frac{\delta_{ij}}{3}\tau_{kk}=-2C\overline{\Delta}^2|\overline{S}|\overline{S}_{ij}+K\left[L_{ij}^m-\frac{\delta_{ij}}{3}L_{kk}^m\right],
\end{equation}
where the first term on the right-hand-side (RHS) is due to the eddy-viscosity Smagorinsky
model, while the second term is the modified Leonard stress term $L_{ij}^m$, which is given by
\begin{equation}
L_{ij}^m=\overline{\overline{u}_i\overline{u}_j}-\overline{\overline{u}}_i\overline{\overline{u}}_j.
\end{equation}
The modified Leonard stress is a known quantity from the resolved field. At the test filter level,
the SGS stress can be written as
\begin{equation}
T_{ij}-\frac{\delta_{ij}}{3}T_{kk}=-2C\widetilde{\overline{\Delta}}^2|\widetilde{\overline{S}}|\widetilde{\overline{S}}_{ij}+K\left[L_{ij}^t-\frac{\delta_{ij}}{3}L_{kk}^t\right].
\end{equation}
Using Germano's identity (1991)~\cite{germano91}, we obtain
\begin{equation}
L_{ij}=T_{ij}-\widetilde{\tau}_{ij},
\end{equation}
which leads to
\begin{equation}
L_{ij}-\frac{\delta_{ij}}{3}L_{kk}=-2C\overline{\Delta}^2M_{ij}+K\left[H_{ij}-\frac{\delta_{ij}}{3}H_{kk}\right],
\end{equation}
where
\begin{eqnarray}
L_{ij}&=&\widetilde{\overline{u}_i\overline{u}_j}-\widetilde{\overline{u}}_i\widetilde{\overline{u}}_j,\nonumber\\
H_{ij}&=&\widetilde{\overline{\overline{u}}_i\overline{\overline{u}}_j}-\widetilde{\overline{\overline{u}}}_i\widetilde{\overline{\overline{u}}}_j,\nonumber\\
M_{ij}&=&\alpha^2|\widetilde{\overline{S}}|\widetilde{\overline{S}}_{ij}-\widetilde{\overline{S}|\overline{S}_{ij}}.\nonumber
\end{eqnarray}

Here, all the above tensors, $L_{ij}$ , $H_{ij}$ and $M_{ij}$, are known from the resolved field and $\alpha=\widetilde{\overline{\Delta}}/\overline{\Delta}$ is
the ratio of test-filter and grid-filter sizes. To obtain the constants $C$ and $K$, the least
squares technique in the context of dynamic SGS modeling as proposed by Lilly (1992)~\cite{lilly92} is employed. Thus,
the constants are obtained by minimizing the following function with respect to these
parameters:
\begin{equation}
Q=\left\{L_{ij}-\frac{\delta_{ij}}{3}L_{kk}+2C\overline{\Delta}^2M_{ij}-K\left[H_{ij}-\frac{\delta_{ij}}{3}H_{kk}\right]
   \right\}^2.
\end{equation}

With $\partial Q/\partial C = 0$ and $\partial Q/\partial K = 0$, we obtain
\begin{eqnarray}
K&=&\frac{ab/d-m}{a^2/d-n},\\
C&=&\frac{aK-b}{2\overline{\Delta}^2d},
\end{eqnarray}
where $a=\left[H_{ij}-\frac{\delta_{ij}}{3}H_{kk}\right]M_{ij}$, $b=\left[L_{ij}-\frac{\delta_{ij}}{3}L_{kk}\right]M_{ij}$,
$d=M_{ij}M_{ij}$, $m=\left[L_{ij}-\frac{\delta_{ij}}{3}L_{kk}\right]\left[H_{ij}-\frac{\delta_{ij}}{3}H_{kk}\right]$, and
$n=\left[H_{ij}-\frac{\delta_{ij}}{3}H_{kk}\right]\left[H_{ij}-\frac{\delta_{ij}}{3}H_{kk}\right]$.

It may be noted that when DMM is employed, with $K=1$, the Smagorinsky coefficient becomes
\begin{equation}
C=\frac{aK-b}{2\overline{\Delta}^2d}=\frac{\left[\left(L_{ij}-\frac{\delta_{ij}}{3}L_{kk}\right)-\left(H_{ij}-\frac{\delta_{ij}}{3}H_{kk}\right)\right]}{2\overline{\Delta}^2M_{ij}M_{ij}}.
\end{equation}

Once the local values of the coefficients $C$ and $K$ are known, the SGS stress can be
incorporated within the framework of lattice Boltzmann method (LBM) as follows: The first part
of the SGS stress in DTM provides the eddy viscosity
\begin{equation}
\nu_t=C\overline{\Delta}^2|\overline{S}|
\end{equation}
which can be added to the molecular viscosity $\nu_0$, obtained from the statement of the problem, to
yield the total viscosity $\nu$ , i.e., $\nu=\nu_0+\nu_t$ can then be used to compute relaxation times
$(s_{10},s_{12},s_{14},s_{15},s_{16})$ in the MRT-LBE, as discussed in the main text of this paper for the DSM.
The second part of the SGS stress can be introduced as forcing terms in the MRT-LBE. Thus, the
modified Leonard stress can be written as the following Cartesian components of the effective
force:
\begin{eqnarray}
F_x^{mL}&=&-\left[\partial_x(KL_{xx}^m)+\partial_y(KL_{xy}^m)+\partial_z(KL_{xz}^m)\right],\\
F_y^{mL}&=&-\left[\partial_x(KL_{yx}^m)+\partial_y(KL_{yy}^m)+\partial_z(KL_{yz}^m)\right],\\
F_z^{mL}&=&-\left[\partial_x(KL_{zx}^m)+\partial_y(KL_{zy}^m)+\partial_z(KL_{zz}^m)\right].
\end{eqnarray}
The spatial derivatives needed in the forcing terms can be evaluated by employing isotropic
spatial discrete derivative operators based on lattice stencils. Such forcing terms $\overrightarrow{F}^{mL}$, where they
are added to $\overrightarrow{F}$ enter the MRT-LBE through the source terms in the moment space $\widehat{S}_{\alpha}$,
defined at the end of Appendix~\ref{app:momentcomponents}.

It may be noted that the grid-filtered fields can be obtained directly from the solution of
the MRT-LBE. On the other hand, the double-grid filtered $\overline{\overline{\phi}}$ as well as
the test-filtered value $\widehat{\overline{\phi}}$, required in the DMM as well as the DSM,
can be obtained from the repeated application of trapezoidal discrete filters for the grid
and test volumes, respectively, successively for each spatial dimension~\cite{zang93}.
Thus, at grid node $(i,j,k)$, for the double-grid filter:
\begin{eqnarray}
\overline{\phi}_{(i,j,k)}^{*}&=&1/8\left(\overline{\phi}_{(i+1,j,k)}+6\overline{\phi}_{(i,j,k)}+\overline{\phi}_{(i-1,j,k)}\right),\nonumber\\
\overline{\phi}_{(i,j,k)}^{**}&=&1/8\left(\overline{\phi}_{(i,j+1,k)}^{*}+6\overline{\phi}_{(i,j,k)}^{*}+\overline{\phi}_{(i,j-1,k)}^{*}\right),\nonumber\\
\overline{\overline{\phi}}_{(i,j,k)}&=&1/8\left(\overline{\phi}_{(i,j,k+1)}^{**}+6\overline{\phi}_{(i,j,k)}^{**}+\overline{\phi}_{(i,j,k-1)}^{**}\right),\nonumber
\end{eqnarray}
and for the test-filter:
\begin{eqnarray}
\overline{\phi}_{(i,j,k)}^{*}&=&1/4\left(\overline{\phi}_{(i+1,j,k)}+2\overline{\phi}_{(i,j,k)}+\overline{\phi}_{(i-1,j,k)}\right),\nonumber\\
\overline{\phi}_{(i,j,k)}^{**}&=&1/4\left(\overline{\phi}_{(i,j+1,k)}^{*}+2\overline{\phi}_{(i,j,k)}^{*}+\overline{\phi}_{(i,j-1,k)}^{*}\right),\nonumber\\
\widetilde{\overline{\phi}}_{(i,j,k)}&=&1/4\left(\overline{\phi}_{(i,j,k+1)}^{**}+2\overline{\phi}_{(i,j,k)}^{**}+\overline{\phi}_{(i,j,k-1)}^{**}\right).\nonumber
\end{eqnarray}
It may be noted that implementation of this approach in the MRT-LBE within the context of multiblock
grids would require careful consideration of information exchange between different grid
levels in the grid transition regions. Implementation and assessment of dynamic scale-similarity based SGS models in the GLBE framework are subjects of future investigations.



\end{document}